\documentclass[twocolumn]{aastex631}
\usepackage{ragged2e}

\usepackage{siunitx}
\usepackage{booktabs}
\usepackage{graphicx}

\usepackage{xspace}
\usepackage{apjfonts}
\usepackage{float}
\usepackage{hyperref}
\definecolor{LightCyan}{rgb}{0.88,1,1}
\usepackage{color, colortbl}
\newcommand{\spocsn}{\ensuremath{27_{-8}^{+10} \%}}
\newcommand{\spocmn}{\ensuremath{35_{-10}^{+13} \%}}
\newcommand{\qlpsn}{\ensuremath{13^{+3.9}_{-4.9}}\%}
\newcommand{\qlpmn}{\ensuremath{22^{+8.6}_{-6.8}}\%}

\newcommand{\spocor}{\ensuremath{61^{+14}_{-13}\%}}
\newcommand{\qlpor}{\ensuremath{33^{+10}_{-7.3}\%}}

\newcommand\vsini{\ifmmode{v\sin{i_\star}}\else $v\sin{i_\star}$\fi}

\newcommand\sini{\ifmmode{\sin{i_\star}}\else $\sin{i_\star}$\fi}

\newcommand{\Kepler}{{\it Kepler}}

\newcommand\mysim{\mathord{\sim}}

\newcommand{\cfa}{Center for Astrophysics \textbar{} Harvard \& Smithsonian, 60 Garden Street, Cambridge, MA 02138, USA}

\newcommand{\MIT}{Department of Physics and Kavli Institute for Astrophysics and Space Research, Massachusetts Institute of Technology, Cambridge, MA 02139, USA}

\newcommand{\usq}{Centre for Astrophysics, University of Southern Queensland, West Street, Toowoomba, QLD 4350, Australia}

\usepackage{amsmath}

\shorttitle{Young Planet Occurrence Rates}
\shortauthors{Vach et al.}

\graphicspath{{./}{figures/}}
\begin{document}
\title{The occurrence of small, short-period planets younger than 200 Myr with \textit{TESS}}


\author[0000-0001-9158-9276]{Sydney Vach}
\affiliation{\usq}

\author[0000-0002-4891-3517]{George Zhou}\affiliation{\usq} 

\author[0000-0003-0918-7484]{Chelsea X. Huang}\affiliation{\usq}

\author[0000-0001-7615-6798]{James G. Rogers}\affiliation{Department of Earth, Planetary, and Space Sciences, The University of California, Los Angeles, 595 Charles E. Young Drive East, Los Angeles, CA 90095, USA}

\author[0000-0002-0514-5538]{L. G. Bouma} 
\altaffiliation{51 Pegasi b Fellow}
\affiliation{Cahill Center for Astrophysics, California Institute of Technology, Pasadena, CA 91125, USA}

\author[0000-0001-7371-2832]{Stephanie T. Douglas}  
\affiliation{Department of Physics, Lafayette College, 730 High St., Easton, PA 18042, USA}

\author{Michelle Kunimoto}
\affiliation{\MIT}

\author[0000-0003-3654-1602]{Andrew W. Mann}
\affiliation{Department of Physics and Astronomy, The University of North Carolina at Chapel Hill, Chapel Hill, NC 27599, USA}

\author[0000-0002-8399-472X]{Madyson G. Barber}
\altaffiliation{NSF Graduate Research Fellow}
\affiliation{Department of Physics and Astronomy, The University of North Carolina at Chapel Hill, Chapel Hill, NC 27599, USA}

\author[0000-0002-8964-8377]{Samuel N. Quinn} 
\affiliation{\cfa}

\author[0000-0001-9911-7388]{David W. Latham}
\affiliation{\cfa}

\author[0000-0001-6637-5401]{Allyson Bieryla}
\affiliation{\cfa}
\affiliation{\usq}

\author[0000-0001-6588-9574]{Karen Collins}
\affiliation{\cfa}

\begin{abstract}
Within the first few hundreds of millions of years, many physical processes sculpt the eventual properties of young planets. NASA's Transiting Exoplanet Survey Satellite (TESS) mission has surveyed young stellar associations across the entire sky for transiting planets, providing glimpses into the various stages of planetary evolution. 
Using our own detection pipeline, we search a magnitude-limited sample of 7219 young stars ($\lesssim 200$ Myr) observed in the first 4 yr of TESS for small ($2-8\,R_\oplus$), short period ($1.6-20$ days) transiting planets. The completeness of our survey is characterized by a series of injection and recovery simulations. 
Our analysis of TESS 2 minute cadence and Full Frame Image (FFI) light curves recover all known TESS Objects of Interest (TOIs), as well as four new planet candidates not previously identified as TOIs. We derive an occurrence rate of \spocmn\ for mini-Neptunes and \spocsn\ for super-Neptunes from the 2 minute cadence data, and \qlpmn\ for mini-Neptunes and \qlpsn\ for super-Neptunes from the FFI data. 
To independently validate our results, we compare our survey yield with the predicted planet yield assuming Kepler planet statistics. 
We consistently find a mild increase in the occurrence of super-Neptunes and a significant increase in the occurrence of Neptune-sized planets with orbital periods of $6.2-12$\,days when compared to their mature counterparts. The young planet distribution from our study is most consistent with evolution models describing the early contraction of hydrogen-dominated atmospheres undergoing atmospheric escape and inconsistent with heavier atmosphere models offering only mild radial contraction early on.   
\end{abstract} 


\section{Introduction} \label{sec:intro}
During the first $\mysim$100 Myr postformation, many evolutionary processes sculpt the eventual properties and architectures of young planetary systems. Planet-disk \citep[][ $\lesssim$ 10 Myr]{Lin:1989,Ida:2004} and planet-planet and planet-planetesimal ($\lesssim$ 100 Myr) interactions \citep{Chatterjee:2008, Nagasawa:2008, Nagasawa:2011, Schlichting:2015} can cause perturbations to planetary orbits, shaping the eventual system stability and architecture. Further, the evolution of the host star and its interactions with evolving planets ($\lesssim 1$ Gyr) can shape their physical properties \citep[e.g.,][]{Lammer:2003,Jackson:2012,owen:2013,ginzburg2018}. 

Population studies of mature planetary systems have unveiled scarcities and overdensities within the period-radius distribution of close-in small planets, motivating investigations into the mechanisms sculpting this distribution \citep[e.g.][]{Beauge:2013,Mazeh:2016,Fulton:2017, Cloutier:2020,Hardegree-Ullman:2020}. Thermal contraction, paired with atmospheric erosion and mass loss driven by photoevaporation \citep{Lopez:2013,owen:2013} and core-powered mass loss \citep{ginzburg2018, Gupta:2019} are thought to be the dominant processes. Atmospheric erosion driven by photoevaporation resulting from high energy X-ray and UV (XUV) flux is thought to be most active within the first $\mysim100$ Myr, whereas core-powered mass loss is thought to be a more subtle process predicted to occur over a span of $\mysim1$ Gyr. However, recent studies have illustrated that the exact timescales and interplay between these two mechanisms are not yet well understood and require further study \citep{Owen:2023}. As a result of these processes, the demographics of the close-in, small planet population most heavily impacted by these evolutionary processes should vary as a function of age. Kepler \citep{Kepler} and K2 \citep{k2} identified a number of young planets \citep[e.g.,][]{David:2016,Mann2016,Mann2018, Rizzuto2018, Vanderburg2018, David:2019, Barber:2022, Bouma:2022}. These surveys provided the first glimpse into the temporal evolution that small planets undergo early in their lives. 

Since the launch of NASA's Transiting Exoplanet Survey Satellite \citep[TESS;][]{Ricker2016} in 2018, we have now been able to observe $\mysim95\%$ of the sky. This has allowed us to gather precise space-based photometry for stars located in well-aged moving groups and associations across the entire sky, providing the first opportunity for detailed investigations into the occurrence rates of planets as a function of age. 

TESS has already detected and characterized tens of young planetary systems \citep[e.g.,][]{Newton2019, Battley:2020,Plavchan2020,zhou:2022, Wood:2023}, providing initial insights into the differences between young planetary systems and their mature counterparts. Initial investigations into the population of young planets observed by TESS span a wider age range \citep{Fernandes:2023}, targeting clusters which already have confirmed planets \citep{Fernandes:2022}, and explored methods of light-curve extraction that best preserve transit signals in the scenario of heightened photometric variability from stellar activity \citep{PATHOS}.  

In this paper, we present the occurrence rates of close-in ($1.6-20$ days), small ($2-8\,R_\oplus$) planets around stars in nearby moving groups and associations younger than 200 Myr. In Section~\ref{sec:stellar_pop}, we describe our parent stellar sample and our selection criteria. Utilizing the photometry provided by TESS we present calculated rotation periods for our parent stellar population in Section~\ref{sec:stellar_pop}. Our planet detection pipeline, including our detrending method, transit search, vetting process, and four newly identified planet candidates, are presented in Section~\ref{sec:pipeline}. In Section~\ref{sec:injection}, we present our injection and recovery test results in order to probe the completeness of our planet search pipeline. We present our calculated occurrence rates in Section~\ref{sec:or}. We compare our results to the Kepler yield and complete a forward modeling exercise to validate our derived occurrence rates. We present a discussion of our findings in Section~\ref{sec:discussion}, and our conclusions in Section~\ref{sec:conclusions}. 

\section{TESS Observations}\label{sec:obs}

TESS is an all-sky photometric transit planet survey. TESS observations are conducted via four individual cameras covering a $24^\circ\times96^\circ$ area of the sky, with each sector of observations lasting an average of 27 days. TESS samples the light curves of preselected target stars at 2 minute and 20 s cadences via target pixel stamp observations. These data products are processed by the Science Processing Operation Center \citep[SPOC;][]{Jenkins2016} at the NASA Ames Research Center. We make use of the SPOC Simple Aperture Photometry (SAP) for our SPOC planet search. 

Additionally, TESS observes its entire field of view via Full-Frame Images (FFIs). FFIs were taken every 30 minutes in the TESS primary mission and then every 10 minutes in the first extended mission. The FFIs are downlinked and processed by the MIT Quick-Look Pipeline \citep[QLP;][]{qlp2020a, qlp2020b, Kunimoto:2022}. We utilize the QLP default aperture raw flux light curves for our QLP planet search program. 

We make use of the light curves made available through the Mikulski Archive for Space Telescopes (MAST)\footnote{\url{https://mast.stsci.edu/portal/Mashup/Clients/Mast/Portal.html}} as of 2023 February. These include the first 4 yr (Sectors 1-58) of TESS.\footnote{We do not make use of newer 200 s FFI light curves in this analysis as they were made available post-February 2023.}

\begin{figure}
    \centering
    \includegraphics[width=\linewidth]{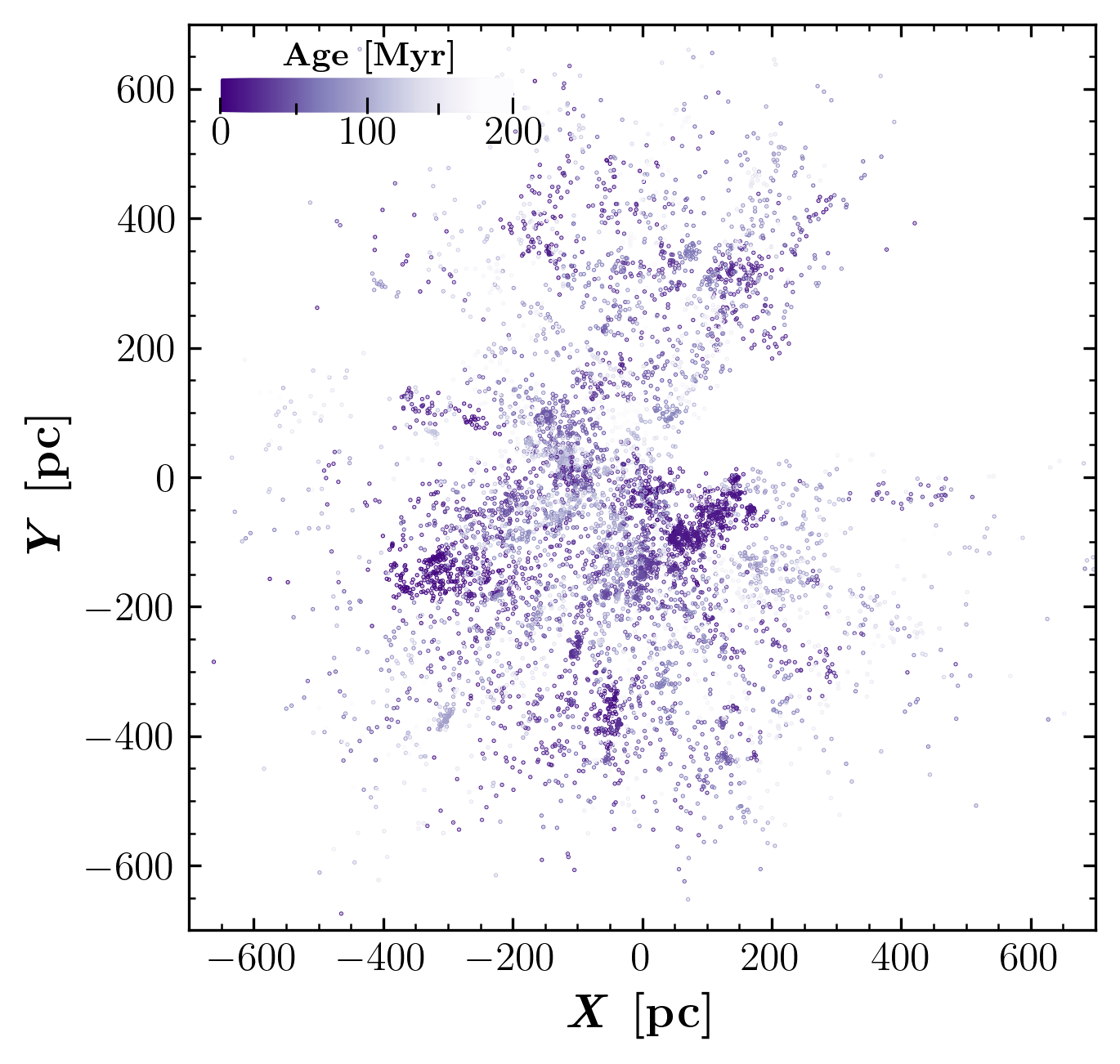}
    \caption{Galactic XY coordinates of the parent stellar sample from Gaia as a function of age (darker colors indicate younger literature ages). }
    \label{fig:XYage}
\end{figure}

\section{Stellar Population}
\label{sec:stellar_pop}


\begin{figure*}[ht!]
    \centering
    \includegraphics[width=\linewidth]{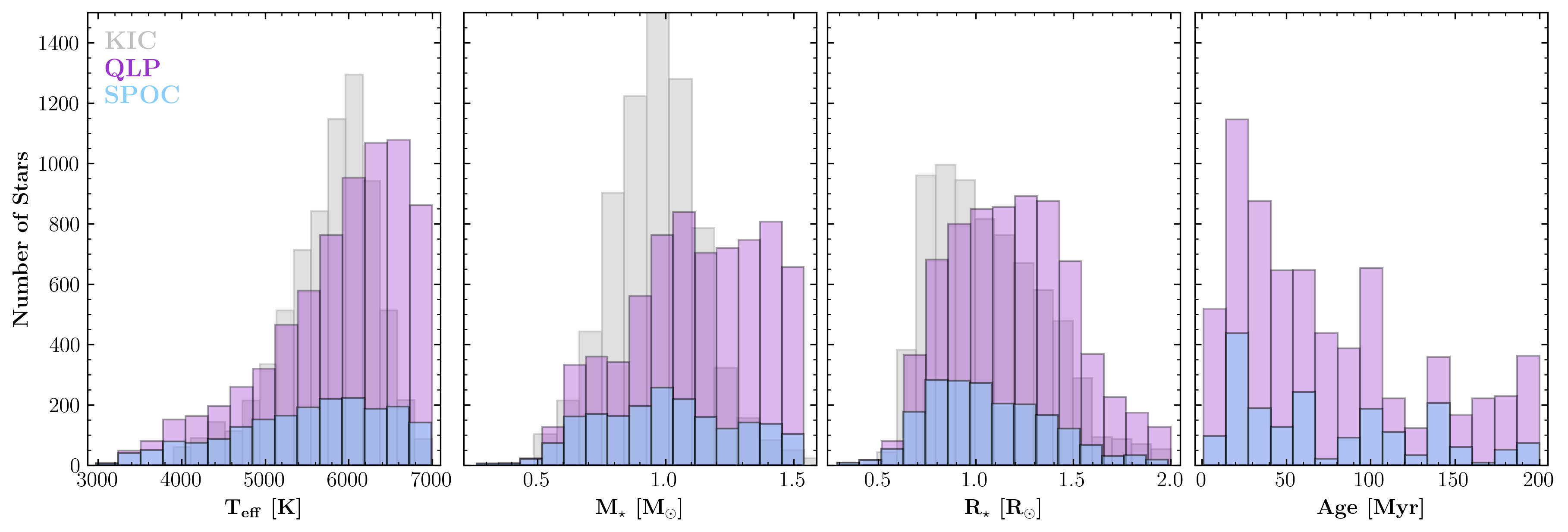}
    \caption{Histograms of the stellar properties of the parent stellar sample in our survey. To select our parent population, we perform a radius cut at $<2\,R_\odot$, a mass cut at $<2\,M_\odot$, a temperature cut at $<7000$\,K, and an age cut at $<200$ Myr. The stellar sample observed by SPOC (1927 stars) is shown in blue, and the stellar sample observed by TESS in the FFIs (7154 stars) is shown in purple. The left panel shows a histogram of the effective temperatures of the parent sample, the middle panel shows a histogram of the stellar masses, and the right panel shows a histogram of the stellar radii \citep{Stassun:2019}. We plot a random sample of 7154 stars of the \Kepler\ Input Catalog \citep{Mathur:2017} in gray for comparison to our parent sample.}
    \label{fig:stellar_pop_hist}
\end{figure*}

ESA's Gaia \citep{Gaia2016,GaiaDR22018, GaiaDR3} mission has provided relative motions, proper motions, and parallaxes for almost all nearby stars, which has revealed previously unknown comoving populations. 

We collate a sample of stars from literature analyses of Gaia kinematics to assemble a population of young stars for our planet search. We make use of associations and clusters assembled from \citet{Gagne:2018}, \citet{Kounkel:2019}, \citet{Ujjwal2020}, and \citet{Moranta:2022}, as well as identified members for well-characterized associations and clusters from \citet{Rebull:2016, Rebull:2018,Rebull:2022}, and \citet{Meingast:2019}. \citet{Gagne:2018} and \citet{Ujjwal2020} provided updated memberships for previously known young stellar associations. \citet{Ujjwal2020} also updated the ages of the associations via Gaia magnitudes and MIST isochrone modeling \citep{Dotter:2016}. \citet{Kounkel:2019} applied unsupervised learning to Gaia DR2 space motions to assemble groups of associated stars, named as Theia groups, and aged via isochrone fits to their Gaia magnitudes. \citet{Moranta:2022} searched for nearby stellar streams within $\mysim200$ pc using $6D$ kinematics from Gaia EDR3, named Crius groups. For newly identified groups, \citet{Moranta:2022} estimate age ranges for identified associations via isochrone fits to the Gaia magnitudes. However, we only adopt the members of Crius groups that are affiliated with previously identified associations and clusters that have well-characterized ages. \citet{Rebull:2016} and \citet{Rebull:2018} use literature photometry and Gaia astrometry to classify previously identified members of the Pleiades cluster and Upper Sco; we make use of stars identified as best/gold members (see \citealt{Rebull:2016} Table 1; \citealt{Rebull:2018} Table 1). \citet{Rebull:2022} select members of Upper Centarus-Lupus (UCL) and Lower Centaurus-Crux (LCC) from the literature, additionally requiring $V$ and $K$ photometry and no source confusion in TESS; we only adopt members identified as gold members (see \citealt{Rebull:2022} Table 2). \citet{Meingast:2019} used 6D Gaia kinematics to identify overdensities in velocity space, indicative of nearby stellar streams. We make use of the members identified to be part of Pisces-Eridanus stellar stream \citep{Curtis2019b,Meingast:2019}. The assigned associations for each source star are listed in Table~\ref{tab:stars}; where a star is listed from more than one source, all references and literature association assignments are noted. Figure~\ref{fig:XYage} shows the Galactic coordinates of our stellar population. 

We concatenate these membership lists and crossmatch them with TICv8 \citep{Stassun:2019}. TICv8 stellar parameter values \citep{Stassun:2019} are adopted for all subsequent analyses. With the exception of planet-hosting stars, we do not propagate the stellar parameter uncertainties for individual stars forward in the occurrence rate analyses. We then select for stars with age estimates $\leqslant200$ Myr, stellar radii $\leqslant 2\,R_\odot$, stellar masses $\leqslant 2\,M_\odot$, effective temperatures $\leqslant 7000$ K, and TESS band magnitudes $\leqslant 12$ (see Figure~\ref{fig:stellar_pop_hist}). This yields 9847 unique stars, with 8804 having been observed by TESS.

\begin{figure}[ht!]
    \centering
    \includegraphics[width=\linewidth]{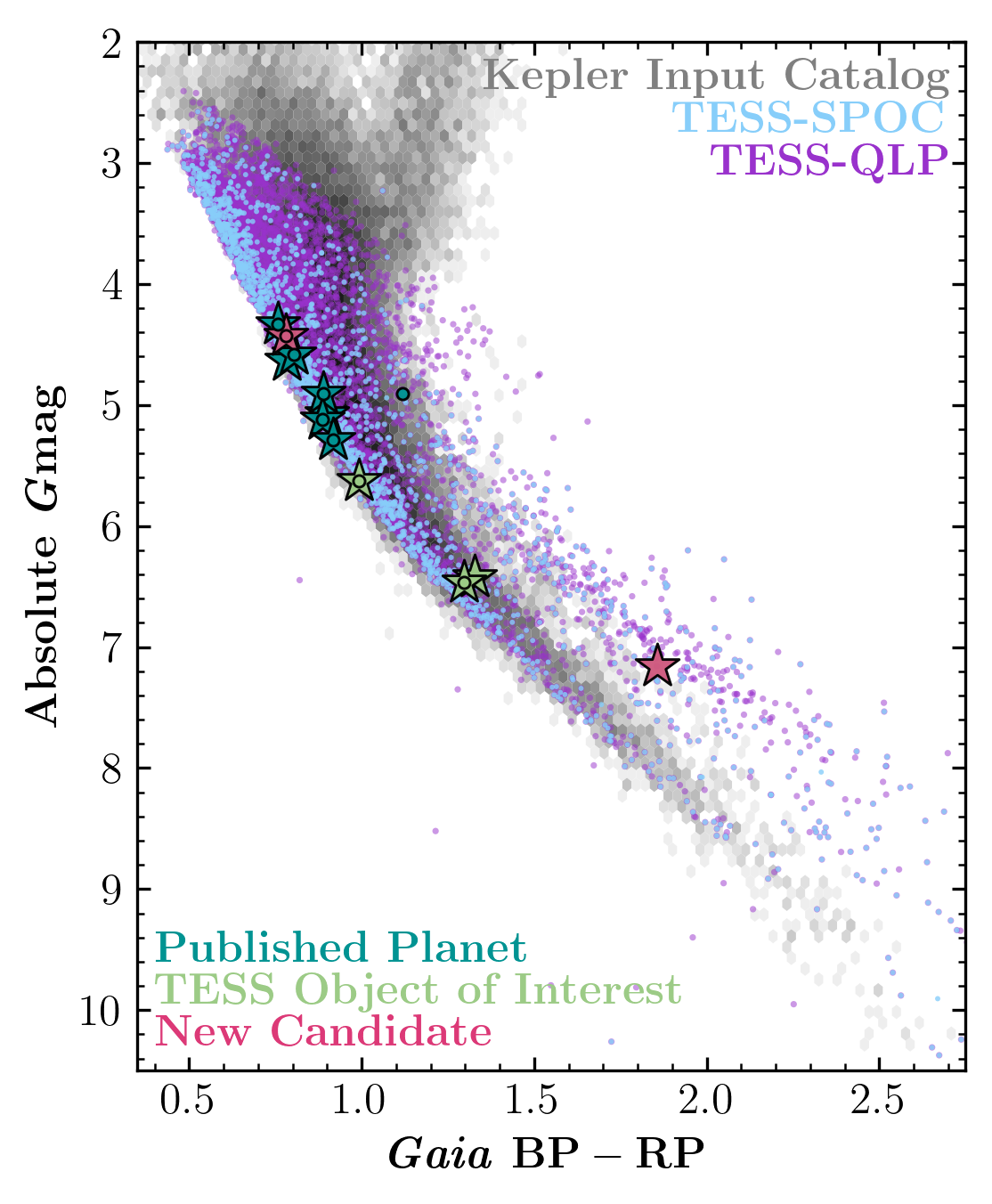}
    \caption{CMD of the stellar population. The comoving stars that have been kinematically aged to be $\lesssim 200$Myr that have been observed by TESS in 2 minute cadence (light blue) and in the TESS FFIs (purple) are plotted over the Kepler Input Catalog \citep[][gray]{Mathur:2017}. Overlaid are the planet hosts identified in our SPOC (circles) and the QLP (stars) surveys. The dark blue are published planetary systems, the green are known TESS Objects of Interest (TOIs), and the pink are new planet candidates identified in this search.}
    \label{fig:cmd}
\end{figure}

Following previous planet occurrence studies with the Kepler sample \citep[e.g.,][]{Berger:2020, Kunimoto2020}, we filter for binaries by removing stars with a Gaia DR3 renormalized unit weight error (RUWE) $>1.4$ \citep{Kervella:2022}. However, a blind RUWE cut removes stars from our sample that have a resolved companion. Therefore, for any star with RUWE $>$ 1.4, we check for a resolved companion within 2'' and a delta magnitude of 4, as the resolved companion explains the high RUWE and does not suggest a possible false positive \citep{El-Badry:2021}. If this criterion is matched, the high-RUWE star is included in our parent population. This yields 7219 stars with TESS observations. Of these, 7154 were observed in the TESS FFIs and 1927 were preselected to have SPOC 2 minute cadence data. Figure~\ref{fig:cmd} shows a CMD of our parent stellar population. 

The population of stars within our SPOC sample is biased toward the proposed science cases for their observations.
As a result, the distributions of stellar properties in our SPOC and QLP parent samples differ. Figure~\ref{fig:stellar_pop_hist} shows our QLP sample is skewed toward brighter, earlier-type stars leading to intrinsic differences in our two surveys' parent populations. 

\begin{table*}[ht!]
    \caption{Parent stellar population}
    \begin{tabular}{lllcclllcll}
\toprule \toprule
TIC & R.A. & Dec. & $T_\mathrm{eff}$ (K) & Parallax (mas) & $R_\star\, (R_\odot)$ & $M_\star\, (M_\odot)$ & TESS$_\mathrm{mag}$ & Age (Myr) & Association & $P_{\mathrm{rot}}$ (days) \\ 
\midrule \midrule
289434416	&315.36223&	50.52834&	6400&	2.11&	1.26&	1.6	&11.3	&1.0&	Theia1	&\nodata \\
454364155	&168.17869&	-76.61802&	4700&	5.23&	0.75&	1.45	&11.2	&1.0&	Theia2	&6.1 \\
454364158	&168.10123&	-76.61838&	5100&	5.16&	0.86&	1.73	&10.5	&1.0&	Theia2	&2.9 \\
& & & & &. . . &  & & & & \\
\bottomrule

\end{tabular}\\

\centering
The full version of this table is made available in the supplementary online material.
    \label{tab:stars}
\end{table*}

Our target list is a collation of multiple catalogs, some of which encompass the traditional clusters and associations, while others are composed of new associations identified via clustering studies. A fraction of our target list may be contamination stars that are not part of their assigned associations, while others may have ages that are older than our thresholds. For example, \citet{Kounkel:2019} estimate a field star contamination rate of 5-10\% for the Theia group memberships. From the catalogs we used to collate our parent stellar sample, only \citet{Kounkel:2019} estimate an age uncertainty, $\mysim0.15$ dex, which we adopt for our entire parent sample in our forward modeling exercises (see Section~\ref{sec:forwardmodel}). 

To independently verify the contamination rate of our target sample, we measure the rotation periods and variability signatures of our sample with the available TESS photometry. We attempt to derive rotation periods for each star with TESS input catalog effective temperatures $T_\mathrm{eff} < 6500$\,K via a Lomb-Scargle \citep{lomb1976, scargle1982} periodogram over all available TESS long-cadence photometry from the QLP. We only attempt to derive rotation periods for later-type stars in our sample, as they are more likely to host convective envelops and exhibit strong stellar rotational variability at ages up to 200 Myr. The membership contamination rates estimated from the rotational variability of late-type stars are extrapolated and adopted for our entire survey sample (see Section~\ref{sec:cosmoabc}). 
We search for stellar rotation periods between 0.5 and 12 days, as we do not attempt to search for rotational periods longer than that of the TESS orbital period. We note that stars with ages $<200$ Myr should have rotation periods $<10$ days \citep[e.g.][]{2021A&A...645A.144M}, and as such the upper limit cutoff should not bias our sample. The rotation periods are then manually vetted to check if the signals resemble that of (1) rotational variability, (2) binarity, (3) nonrotational variability, or (4) no variability. Example light-curve figures that were used for our manual vetting process are shown in Appendix~\ref{sec:rotation_plots}. We analyzed rotational signatures for a sample of 6402 stars that satisfy our stellar properties thresholds, of which 4878 exhibit variability resembling rotational modulation from our manual vetting. Our measured rotation periods are presented in Table~\ref{tab:stars}.

Rotation-based age estimates are unreliable for the youngest stars we sample, as pre-main-sequence stars that are still contracting exhibit a significant spread in rotation periods for a given age. As such, we do not directly make use of gyrochronological relationships to directly compare against literature-reported isochrone-derived ages. We do make use of the age-rotation relationships to quantify the membership contamination rate for our sample.  

We make use of the \citet{bouma2023} gyrochronology relationship to convert rotation periods to age posteriors on a star-to-star basis. This is only performed for a subsample of stars that exhibit rotational variability and are not identified to be binaries based on their light curves and Gaia stellar parameter flags. We find that 9\% of stars have gyrochronology-based posteriors with $1\sigma$ age upper limits $>200$ Myr, making them inconsistent with the bounds of our sample selection. In addition, late-type young stars should exhibit significant stellar activity due to spot-induced modulations in their light curves. Our manual vetting flagged 30\% of G and K stars as not exhibiting significant photometric variations. As such, we conservatively adopt a membership contamination rate ranging over 10-30\% for our parent sample in this study.

\begin{figure}[ht!]
    \centering
    \includegraphics[width=\linewidth]{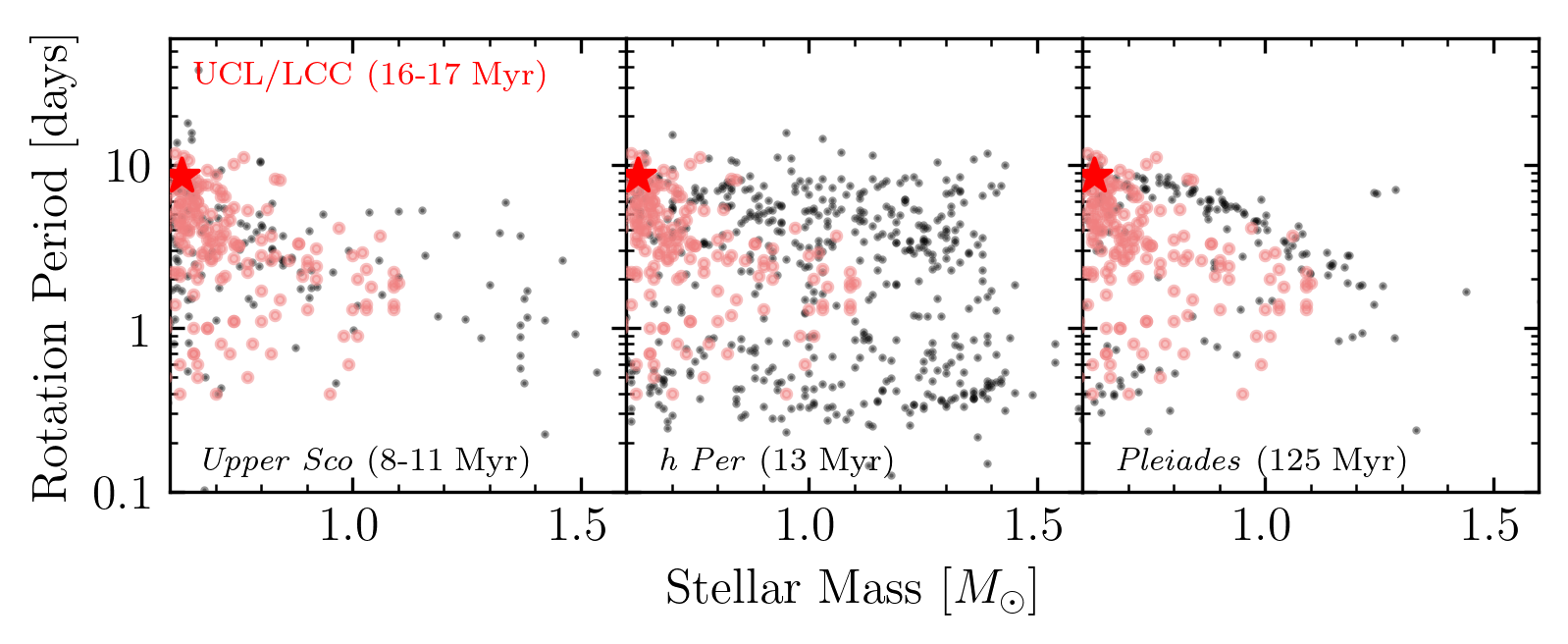}\\
    \includegraphics[width=\linewidth]{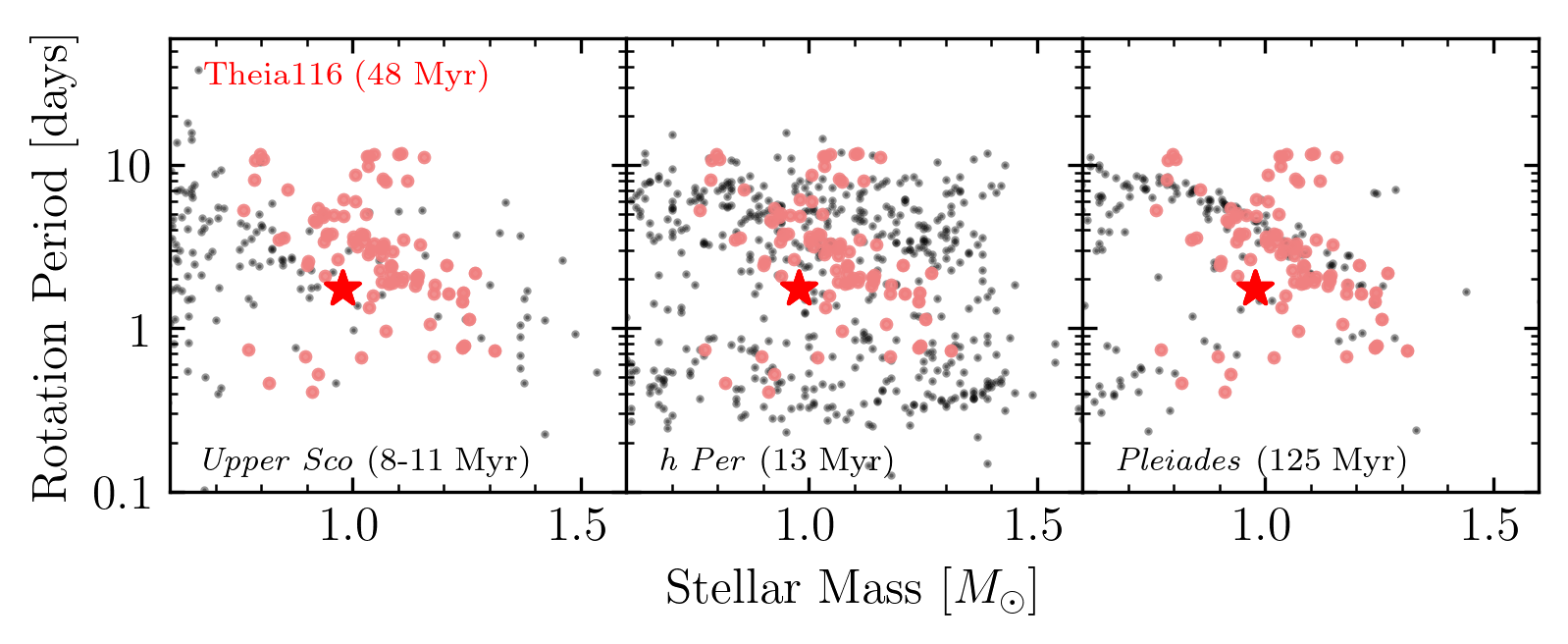}\\
    \includegraphics[width=\linewidth]{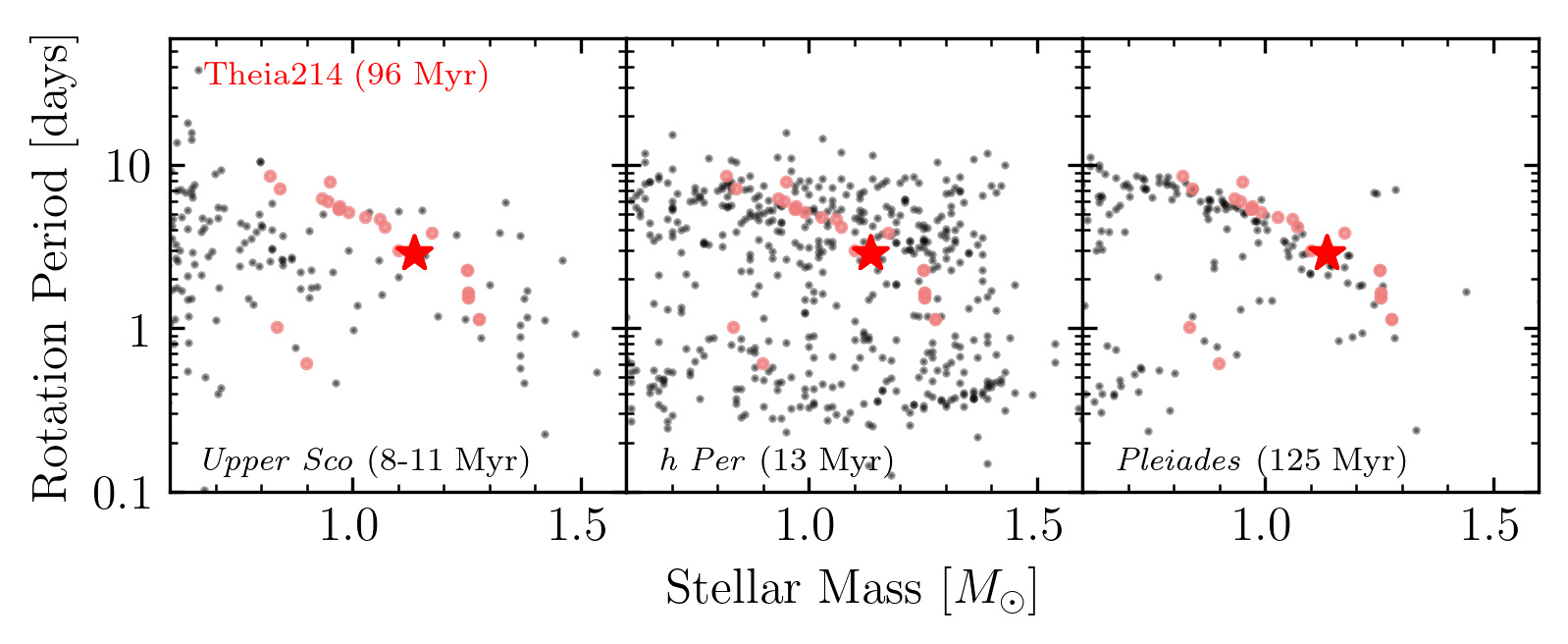}\\
    \caption{Stellar rotation period plotted as a function of stellar mass for stars identified to host planet candidates (red star) and the members of their respective stellar associations (light red): UCL/LCC (upper panel), Theia 116 (middle panel), and Theia 214 (lower panel). We compare the rotation periods of the stellar associations to rotation periods of well-characterized clusters at a variety of stages: Upper Sco \citep[left;][]{Rebull:2018}, h Per \citep[middle;][]{Moraux:2013}, and Pleiades \citep[right;][]{Rebull:2016, Godoy-Rivera:2021}. We compare our derived rotation periods from the TESS QLP light curves to check for youth, and identify possible falsely identified nonmembers.}
    \label{fig:rotation_comparison}
\end{figure}

In addition, we also specifically check for consistency between the literature-reported ages and our gyrochronology-derived ages for the associations that host new planet candidates identified in this study. Figure~\ref{fig:rotation_comparison} shows our measured rotation period distributions of LCC, Theia 116, and Theia 214 against those of previously well-characterized populations. These associations host new planet candidates. We find general consistency between their literature ages to the rotation distributions of their members, and that no planet candidates need to be excluded from our occurrence rate due to their parent association being older than 200 Myr.

\section{Planet Detection Pipeline}
\label{sec:pipeline}

We describe in the sections below our transit search routine, as well as a series of checks to analyze the completeness and reliability of our transit search pipeline. We present our recovered planet candidates in Table~\ref{tab:tois}.

\subsection{Detrending Stellar Activity}\label{sec:detrend}
Young stars present additional challenges during the detrending processes of their light curves as a result of heightened activity. Photometric variations on the scale of 1\%-2\% due to rapid stellar rotation, along with stellar flares, require special attention.  

We perform an iterative sigma clipping outlier rejection for major flare rejection and then model and subtract the photometric signal induced by rapid stellar rotation. Our detrending method is as follows.

\begin{itemize}
    \item Perform an initial spline fit of the stellar rotational variability via the routine described in \citet{Vanderburg:2019}.
    \item Identify flare rejection masks via 10 iterations of upward outlier rejection at the $5\sigma$ level with standard deviation calculated by scaling the measured median absolute deviation value. 
    \item Refit stellar variability via the spline model to the raw light curve with the flare rejection masks applied and flatten the resulting light curve (e.g. see Figure~\ref{fig:tic150070085}). 
\end{itemize}

\subsection{Box Least-squares Transit Detection}\label{sec:bls}
With the detrended light curves, we perform a box least-squares \citep[BLS;][]{kovacs:2002} search to identify planetary transits in the TESS light curves. We sample 100,000 periods linearly spaced between 0.5 and 20 days and 500 transit durations linearly sampled between 0.008 times and 0.08 times of each tested orbital period. 

\subsection{Vetting Threshold-crossing Events}\label{sec:tce}

We identify any signal with a calculated signal-to-pink noise ratio \citep[][]{Hartman:2016} $\geqslant$ 8 to be a threshold-crossing event (TCE). For each individual TCE, we perform a visual inspection to vet for false positives resulting from both astrophysical events and detrending artifacts. Each candidate event is inspected to determine if the signal that triggered the TCE is transit shaped. Further, we require a minimum of three transit events in order to determine a period.

We search for the presence of secondary eclipses and differences in the depths between even and odd transits to determine if the transit is from an obvious eclipsing binary (EB). For identified TCEs, we fit for the transit depth of even and odd transits respectively to rule out EB scenarios. To measure the transit depth, we fix all parameters except for transit depth, $\delta$. The transits are modeled using \texttt{batman} \citep{batman}. We explore the best-fit transit depth and posteriors with a Markov Chain Monte Carlo run implemented with \texttt{emcee} \citep{emcee}. We require the difference in our calculated odd and even transit depths to be $<5\sigma$ to not be classified as an EB, though no reported candidate has an odd-even difference $>3\sigma$ in our sample. We visually vet for secondary eclipses by phase folding the detrended TESS light curves at the derived TCE period and transit time. The entire phase-folded light curve is visually inspected for signs of secondary eclipses. Targets exhibiting EB-like secondary eclipses, including those in eccentric orbits, are excluded as planet candidates. 

We check for obvious nearby eclipsing binaries (NEBs) by comparing the pixel-by-pixel light curves centered around the target star to determine if the signal that triggered the TCE is consistent with being on target. We then perform a centroid analysis with the TESS FFIs to create difference images using the \texttt{transit-diffImage}  code base\footnote{\href{https://github.com/stevepur/transit-diffImage}{https://github.com/stevepur/transit-diffImage}} following the techniques described in \citet{Bryson:2013}. For each candidate in each sector, we produce a difference image between the in-transit and out-of-transit pixel stamps. We calculate the centroid of the target star via the TESS pixel response function fit to the difference image. The calculated centroid is compared against the expected centroid of the target star based on a catalog projection. Some offset in the difference image is expected if nearby stars are present, as the centroid naturally shifts away from the dimming star during transit. We flag any centroid offset $> 1$ pixel for visual inspection to determine if the high centroid offset is due to an astrophysical false positive.

Remaining candidates were then checked for planetary multiplicity. We masked the initial signal that triggered the TCE in the nondetrended light curves and then repeated the detrending and transit search (see Sections~\ref{sec:detrend} and~\ref{sec:bls}). Any signal that pass our S/N criteria is visually vetted in the same method as mentioned above. We additionally check for aliasing of the original signal recovered. All remaining candidates are compared to the list of TOIs and known planets. 

\begin{figure}[ht!]
    \centering
    \includegraphics[width=\linewidth]{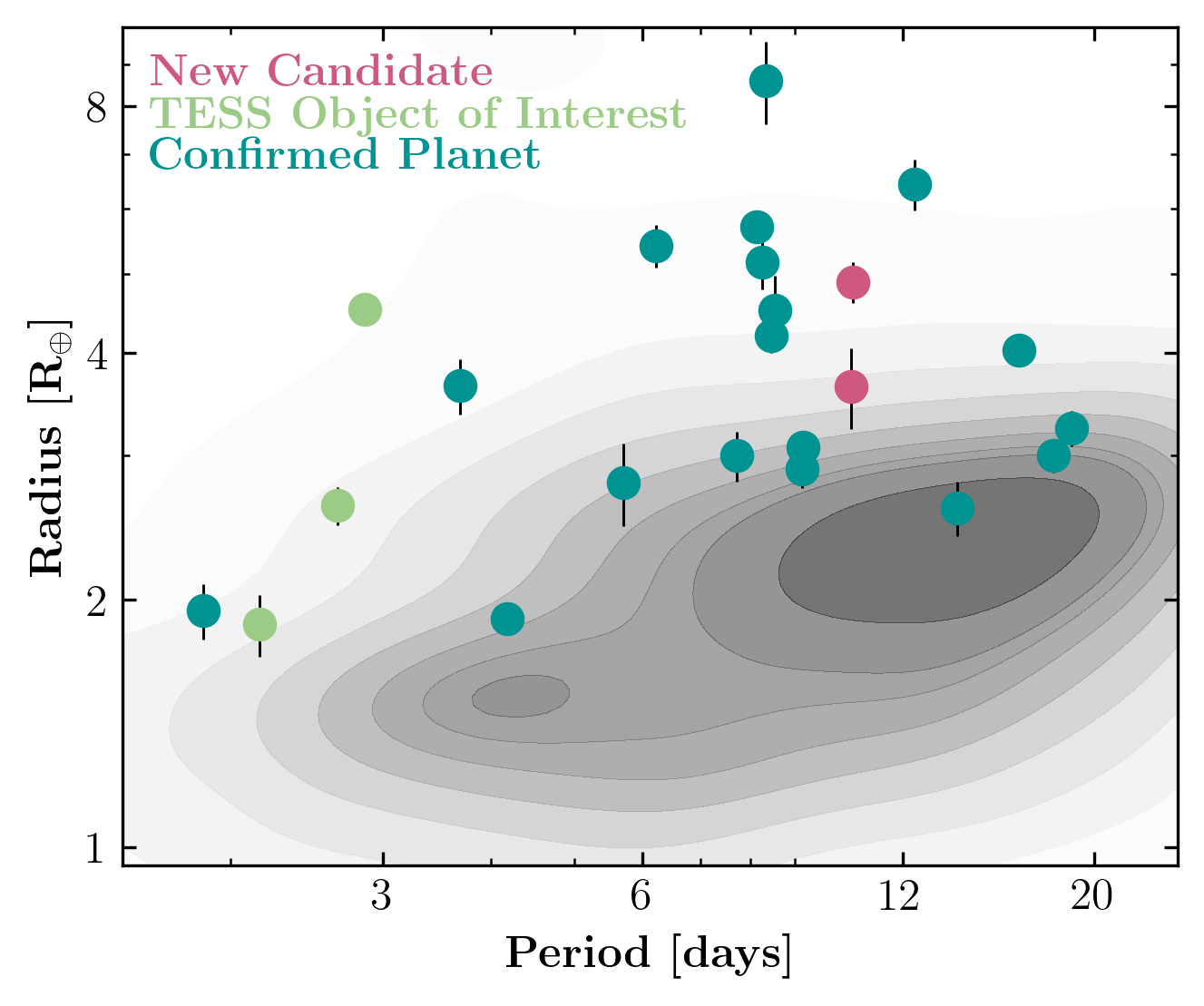}
    \caption{Period-radius distribution of the recovered confirmed planets and planet candidates in our SPOC and QLP surveys. Published planets are shown in blue, planet candidates that are known TOIs are shown in green, and planet candidates newly identified in this work are shown in pink. We plot the contour distribution of the known transiting exoplanet population in gray.}
    \label{fig:rad_per}
\end{figure}

Our stellar sample contained 26 TOIs (see Table~\ref{tab:tois}), 22 of which were within our search parameters (0.5 days $\leqslant\,P\leqslant$ 20 days and $1\,R_\oplus \leqslant\, R_P \leqslant 8\, R_\oplus$), as well as two previously previously confirmed transiting planets from K2. Of these, 16 have both FFI light curves and 2 minute cadence light curves (see Table~\ref{tab:tois}), as well as the two K2 planets. Of the 22 known TOIs within our survey parameters, 20 are either planet candidates or confirmed planets, the remaining two have been ruled as either false alarms or false positives by the TESS Follow-up Observing Program Working Groups (TFOP WG; SG1 and SG2). 

We recovered all known TOIs in our planet search using the SPOC light curves, including 15 planet candidates within our search parameters and the known planets V1298 Tau c and d. Our QLP planet search recovered all 20 TOIs within our parameters but was unable to recover V1298 Tau c and d. Our pipeline identified a signal corresponding to V1298 Tau c, with a signal-to-pink noise ratio of 6.29, which did not trigger a TCE. Upon visual inspection of the FFI light curves for V1298 Tau, the long cadence of the FFIs and the heightened stellar activity subsequently diluted the transit signal in our detection pipeline. 

We identified new planet candidates TIC 88785435.01, TIC 150070085.01, TIC 434398831.01, and TIC 434398831.02 (S. Vach et al. 2024, \textit{in preparation}) in our QLP search. Out of the new planet candidates, only TIC 150070085.01 also has SPOC data and is independently identified by our SPOC search. These candidates are described in detail in Appendix~\ref{newplanets}.

\begin{table*}[ht!]
    \caption{Planets and candidates identified in our search}
    \label{tab:tois}    
\begin{tabular}{rlccccccc}
\toprule \toprule
  TIC &             Name & $R_P$ ($R_\oplus$) & $P$ (days) &    Pipeline$^*$ &  $T_\mathrm{eff}$ (K) &  Age (Myr) & Association & False-positive Rate\\
\midrule \midrule
\multicolumn{9}{l}{\textbf{Confirmed and verified planets}}\\ 
 15756231 &       V1298Tau c$^a$ &      5.16$\pm$0.38 &   8.25 &        SPOC &              5160 &         23 &   Taurus-Auriga Complex & 0\\
 15756231 &       V1298Tau d$^a$&      6.43$\pm$0.46 &   12.41 &        SPOC &              5160 &         23 &     Taurus-Auriga Complex & 0\\
299798795 &      TOI-1224.01$^b$&     1.90$\pm$0.050 &   4.18 & SPOC and QLP &              3390 &        178 &    MELANGE-5 & 0\\
299798795 &      TOI-1224.02$^b$&      3.00$\pm$0.14 &   17.95 & SPOC and QLP &              3390 &        178 &    MELANGE-5 & 0\\
360630575 &      HD 109833 b$^c$ &      2.89$\pm$0.15 &   9.19 &         QLP &              5880 &         26 &     LCC & 0\\
360630575 &      HD 109833 c$^c$&       2.59$\pm$0.2 &   13.90 &         QLP &              5880 &         26 &     LCC & 0\\
384984325 &      TOI-6109.01$^d$&      4.51$\pm$0.46 &   8.54& SPOC and QLP &              5760 &         55 &    Theia133 & 0\\
384984325 &      TOI-6109.02$^d$&      2.78$\pm$0.32 &   5.70 & SPOC and QLP &              5760 &         55 &    Theia133 & 0\\
410214986 &        DS Tuc Ab$^e$&      5.70$\pm$0.17 &   8.14 & SPOC and QLP &              5410 &         40 &     Tuc-Hor & 0\\
441420236 &         AU Mic b$^f$&      4.20$\pm$0.20 &   8.46 & SPOC and QLP &              3590 &         25 &    $\beta$ Pic & 0\\
441420236 &         AU Mic c$^f$&      3.24$\pm$0.16 &   18.86 & SPOC and QLP &              3590 &         25 &    $\beta$ Pic & 0\\
460205581 &        TOI-837 b$^g$    &    8.60$\pm$1.0 &   8.32 & SPOC and QLP &              6100 &         35 &     IC2602 & 0\\
464646604 &      HIP 94235 b$^h$&      3.00$\pm$0.21 &   7.71 & SPOC and QLP &              5990 &        120 &       AB Dor & 0\\
434398831 & TIC 434398831.01$^{i,j}$ &      3.65$\pm$0.28 &   3.69 &         QLP &              5550 &         48 &    Theia116 & 0\\
434398831 & TIC 434398831.02$^{i, j}$ &      5.41$\pm$0.32 &   6.21 &         QLP &              5550 &         48 &    Theia116 & 0\\
257605131 & TOI-451 b$^{k}$ & $1.94\pm0.15$ & 1.86 & SPOC and QLP & 5480 & 120 & Pisces-Eridanus & 0 \\
257605131 & TOI-451 c$^{k}$ & $3.07\pm0.14$ & 9.19 & SPOC and QLP & 5480 & 120 & Pisces-Eridanus & 0 \\
257605131 & TOI-451 d$^{k}$ & $4.03\pm0.15$ & 16.36 & SPOC and QLP & 5480 & 120 & Pisces-Eridanus & 0 \\
\hline
\multicolumn{8}{l}{\textbf{Planet candidates}}\\ 
 46631742 &      TOI-5358.01$^i$ &      2.61$\pm$0.14 &   2.66 & SPOC and QLP &              4640 &        134 &    Theia369 & 0.01126$\pm$0.00088\\
157081737 &      TOI-6095.01$^i$ &      1.87$\pm$0.16 &   2.16 & SPOC and QLP &              5260 &        106 &    Theia301 & 0.1204$\pm$0.0020\\
460950389 &        TOI-6715.01$^i$ &      4.52$\pm$0.19 &   2.86 &         QLP &              4740 &         35 &     Theia92 & 0.00538$\pm$0.00047\\
88785435 & TIC 88785435.01$^i$ & 4.88$\pm$0.28 & 10.51 & QLP & 4000 & 16 & UCL & $0.0947\pm0.0037$ \\
150070085 & TIC 150070085.01$^i$ &      3.64$\pm$0.41 &   10.47 & SPOC and QLP &              6070 &         96 &    Theia214 & 0.0312$\pm$0.0067\\
\hline
\multicolumn{8}{l}{\textbf{Planets and planet candidates not included in occurrence rate calculation}}\\ 

36332984 & TIC 36332984.01$^i$& 4.40 & 6.33 & QLP & 5400& 84 & Theia 217 & \nodata \\
47720259 & TOI-2519.01$^\alpha$& 2.29 & 6.96& QLP & 4740 & 94 & Theia 204 & \nodata\\
64837857 & TOI-6550.01 $^n$ & 14.1 & 20.3 & SPOC and QLP & 6740 & 43 & Theia 117 & \nodata\\
86951294 & TOI-4596.01$^\alpha$& 2.72 & 4.12 & SPOC and QLP & 5600& 55 & Theia 133 & \nodata\\
166527623 & HIP 67522 b $^{l, \beta}$& 10.1 & 6.96 & SPOC and QLP & 5770 & 17 & Sco-Cen & \nodata\\
238395674 & TIC 238395674.01$^i$& -- & -- & QLP & 5700& 22 & Theia 12 & \nodata\\
238395674 & TIC 238395674.02$^i$& -- & -- & QLP & 5700& 22 & Theia 12 & \nodata\\
259606227 & TOI-6555.01$^{i, \gamma}$ & 3.77 & 10.1 & QLP & 6800 & 172 & Theia 436 & \nodata\\
441546821 & HD 114082 b$^{m, \beta}$ & 11.2& -- & SPOC and QLP & 6610 & 16 &LCC &\nodata\\
\hline
\multicolumn{8}{l}{\textbf{TOIs not included in occurrence rate calculation}}\\ 

 14091633&TOI-447.01$^\delta$ & 23.5  &5.52 & SPOC and QLP & 6430  & 35 & Crius 164 & \nodata\\
 151284882&TOI-2595.01$^\delta$ & 3.32 & 3.81 & QLP & 6030 & 48 & Theia 116 & \nodata\\
  117689799&TOI-3504.01$^\delta$ & 14.2 & 2.16 & SPOC and QLP & 5860  & 99 & Theia 254 & \nodata\\
    133505138&TOI-4359.01$^\epsilon$ & 2.47 & 2.436 & SPOC and QLP & 6530  & 91 & Theia 215 & \nodata\\
 164150539&TOI-1372.01$^\zeta$ & 8.22 & 6.16 & SPOC and QLP & 5770  & 100 & Theia 312 & \nodata\\
  294500964&TOI-2550.01$^\zeta$& 10.5 & 9.07 & SPOC and QLP  & 5740  & 83 & Theia 211 & \nodata\\
  457939414&TOI-1990.01$^\eta$ & 11.8 & 3.78 & SPOC and QLP & 6080  & 35 & Theia 92& \nodata \\
\bottomrule

\end{tabular}\\

$(a)$ \cite{David:2019}; $(b)$ P. Thao et al. (2024, \textit{in preparation}); $(c)$ \cite{Wood:2023}; $(d)$ A. Dattilo et al. (2024; \textit{in preparation}); $(e)$ \cite{Newton2019}; $(f)$ \cite{Plavchan2020}; $(g)$ \cite{Bouma:2020}; $(h)$ \cite{zhou:2022}; $(i)$ This work; $(j)$ S. Vach et al. (2024; \textit{in preparation}); $(k)$ \cite{Newton:2021}; ($l$) \citet{Rizzuto:2020}; ($m$) \citet{Zakhozhay:2022}; 
$(\alpha)$ No signatures of youth in spectroscopic follow-up; ($\beta$) Not in our survey parameters; $(\gamma)$ Low SNR, ambiguous signal-- not a TOI as of February 2023; 
$(\delta)$ Ambiguous Planet Candidate; $(\epsilon)$ Nearby Eclipsing Binary SG2; $(\zeta)$ Nearby Planet Candidate; $(\eta)$ Eclipsing Binary SG2; .\\
$^*$The pipeline column identifies which of our surveys recovered the signal.
\end{table*}

\subsection{False-positive Rate}\label{sec:fprate}

Our planet detection pipeline recovered 14 confirmed planets within our SPOC parent sample, and 16 confirmed planets in the QLP parent sample. For planets with confirmed planet dispositions by the TFOP WGs, we adopt a false-positive rate of zero, as these planets have undergone extensive spectroscopic and photometric follow-up observations to rule out false-positive scenarios. 

Five remaining planet candidates have not received sufficient follow up to be categorized as verified or validated planets. As such, there exists a possibility that they are the result of astrophysical false-positive scenarios, instrumental systematics, stellar variability, detrending artifacts, etc. There are insufficient planet candidates around young stars from which we can derive a secure false-positive rate estimate. 

To attain false-positive estimates, we make use of the publically available \texttt{TRICERATOPS} \citep{Giacalone:2021} code to estimate the probability that a transit signal is planetary in nature or the result of an astrophysical false positive. We explore the impact on the false-positive rates when derived using blend analyses codes, including \texttt{TRICERATOPS} and \texttt{MOLUSC} \citep{Wood:2021}. The false-positive rates from \texttt{TRICERATOPS} are adopted in our occurrence rate calculations hereafter and are typically 1-5\% for a given candidate. When we adopt \texttt{TRICERATOPS} and \texttt{MOLUSC} together, the false-positive rate approaches 0\% for all candidates, and its impact on the calculated occurrence rates is the same as if we assume all candidates are true planets. This upper-bound case is further explored in Appendix~\ref{sec:check}. We do not attempt to differentiate the false-positive rates between multi- and single-planet systems. We present our calculated false-positive rates in Table~\ref{tab:tois}. 

We adopt a false-positive rate of 0\% for TIC 434398831.01 and .02, which are presented in this work. These are characterized as confirmed planets as space- and ground-based follow up have ruled out any false-positive scenarios (see Appendix~\ref{newplanets}).

We perform a series of checks to illustrate our false-positive rate assumptions do not significantly influence our derived occurrence rates (see Appendix~\ref{sec:check}).

\subsection{Candidates Identified but Not Included in the Occurrence Rate Calculations}

We identify nine planets or planet candidates that are not included in our occurrence rate calculations. These are presented in Table~\ref{tab:tois}. HIP 67522 b \citep[$10.1\,R_\oplus$,][]{Rizzuto:2020}, HD 114082 \citep[single transit, $11.2\,R_\oplus$,][]{Zakhozhay:2022}, and TOI-6550.01 ($14.1\,R_\oplus$, $P = 20.3$\,days) were recovered in both our SPOC and QLP pipelines, but are not included in our occurrence rate calculations as they fall outside the bounds our survey investigates. We note that TOI-2519.01 (QLP only) and TOI-4596.01 (SPOC and QLP), which we recover in our survey, both show no Li features in ground-based follow-up observations. Upon visual inspection, their respective light curves are also not consistent with youth. We therefore exclude them from our occurrence rate calculations as they are likely not members of their respective clusters. TOI-4596 has an additional single transit event, but we are unable to constrain the period.

TOI-6555.01 is excluded as our vetting processes did not find it to be consistent with being on target. Additionally, TOI-6555.01 was not alerted until Sector 65, which was not made available until after 2023 February.

We identify two new additional candidates, TIC 36332984.01 and TIC 238395674.01, which both pass all of our initial vetting procedures, but we are unable to uniquely determine the orbital periods. TIC 238395674 is likely a multiplanet system, where we observe two transits of TIC 238395674.01 and an additional single transit of TIC 238395674.02. Therefore we do not include these candidates in our occurrence rate calculations. 

Additionally, TOI-6715, previously identified as PATHOS 31 \citep{PATHOS}, has both QLP and SPOC data. However, as of 2023 February only the QLP data were available, therefore we only include TOI-6715 in our QLP occurrence rate calculations, but not in our SPOC occurrence rate calculations.

 \subsection{False-positive TESS Objects of Interest Not Included in the Occurrence Rate Calculations}

We identified seven TOIs that failed our own vetting process but were identified as TOIs by QLP and/or SPOC TPS. These TOIs are presented in Table~\ref{tab:tois}. 

Follow-up observations by the TFOP WGs rule five as nearby planet candidates, meaning the signal triggering the TCE is from a planet transiting a nearby star, or ambiguous planet candidates. TFOP WG SG2 identified the remaining two as a NEB and an EB respectively.

\section{Injection Recovery Test}
\label{sec:injection}
We perform a set of transit injection and recovery tests to test the completeness of the transit detection pipeline. We perform 1000 iterations of transit injections into each of the nondetrended 7154 QLP and 1927 SPOC light curves. The injected transits are modeled via \texttt{batman} \citep{batman}, with periods, $P$, and radii, $R_p$, sampled from log-uniform distributions ranging from 0.5 days to 20 days, and 1$\,R_\oplus$ to $10\,R_\oplus$, respectively, impact parameters, $b$, selected from a uniform distribution, between 0 and 0.8, and a transit time, $T_0$, uniformly selected from a random phase. The models are resampled to 2, 10, and 30 minute cadences, and multiplied into the nondetrended QLP and/or SPOC light curves for each target. The injected light curve is then run through our detection pipeline as described in Section~\ref{sec:pipeline}.  

An injected planet is recovered if the BLS signal-to-pink noise ratio threshold is met and if the output ephemeris matches that of the injection. As with our planet detection pipeline, we require a TCE to have a signal-to-pink noise ratio $\geqslant 8$. The detected TCE is also required to have its detected transit time and period from the BLS to match the injected time and period with a significance of $\Delta \sigma_P \geqslant 2.5$ and $\Delta \sigma_T \geqslant  2$ (see \citet{Coughlin:2014} for definitions of $\Delta \sigma_P$ and $\Delta \sigma_T$). 

\begin{figure*}
    \centering
    \includegraphics[width=0.49\linewidth]{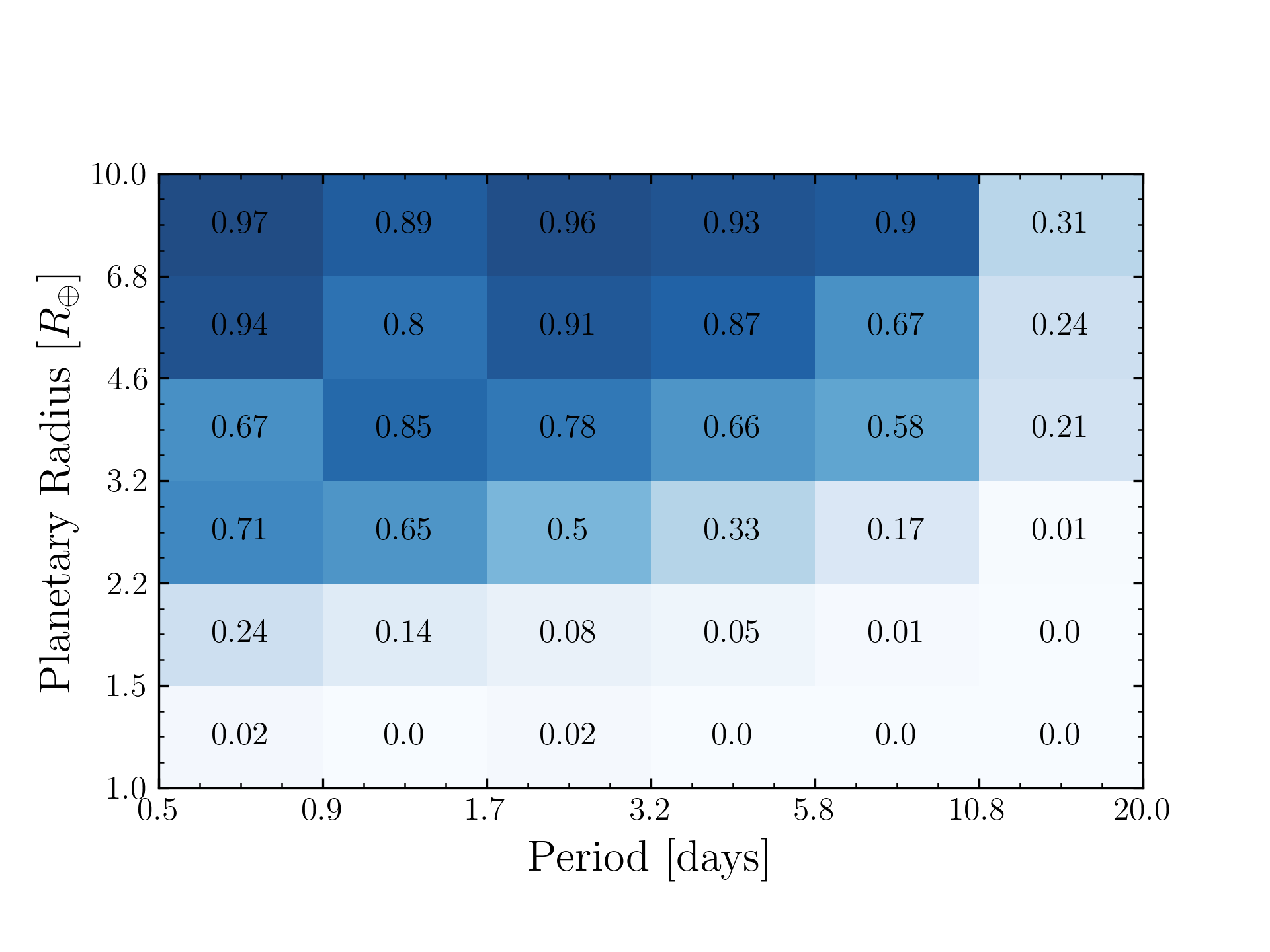}
    \includegraphics[width=0.49\linewidth]{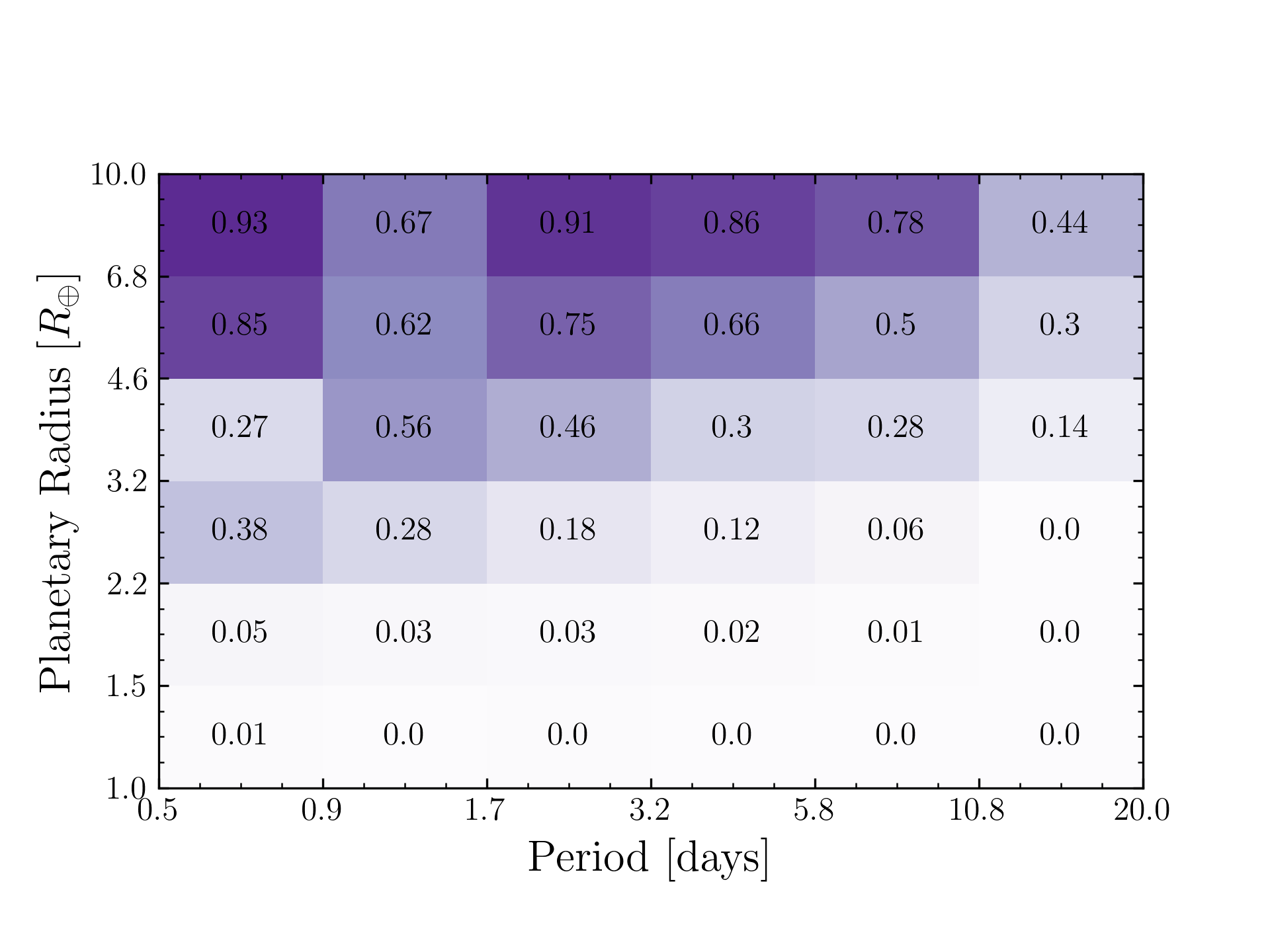}
    \caption{Recoverability maps from a series of injection and recovery tests performed on each star in our SPOC (left) and QLP (right) surveys. We inject 1000 simulated planet signals into the light curves of every star within our survey sampling orbital periods between 0.5 and 20 days, and radii between 1 and 10$\,R_\oplus$. The darker shades indicate a higher recoverability rate. }
    \label{fig:recovery_maps}
\end{figure*}

We present our derived injection and recovery maps in Figure~\ref{fig:recovery_maps}. These maps present our full series of injection and recovery tests summed across the entire parent stellar population of QLP and SPOC respectively. Our recoverability of larger planets ($\gtrsim$5$\,R_\oplus$) is consistent across the two surveys, but we see a sharper decline in recoverability in the region between 2 and 4$\,R_\oplus$ for QLP compared to SPOC. This is consistent with our expectations due to the difference in stellar populations between the two samples. The on-average larger stellar radii in the QLP sample lead to more difficulties in recovering smaller planets. 

We also investigate our susceptibility to false alarms with an inverted light-curve test, as described in Appendix~\ref{sec:inverted}, finding our manual vetting to be sufficient in removing such cases.

\section{Occurrence Rate Calculations}\label{sec:or}

\subsection{Forward Modeling}\label{sec:forwardmodel}

We perform a forward modeling test to compare the demographics of planets around young stars against that of the Kepler distribution. We synthesize a population of planets using the Kepler statistics as presented in \citet{Kunimoto2020} and compute the expected number of recovered planets as per our survey efficiency. The forward modeling process allows us to compare the expected characteristics of the planet population against those that we identify from the young stellar sample. 

\begin{figure*}[ht!]
    \centering
    \includegraphics[width=0.49\linewidth]{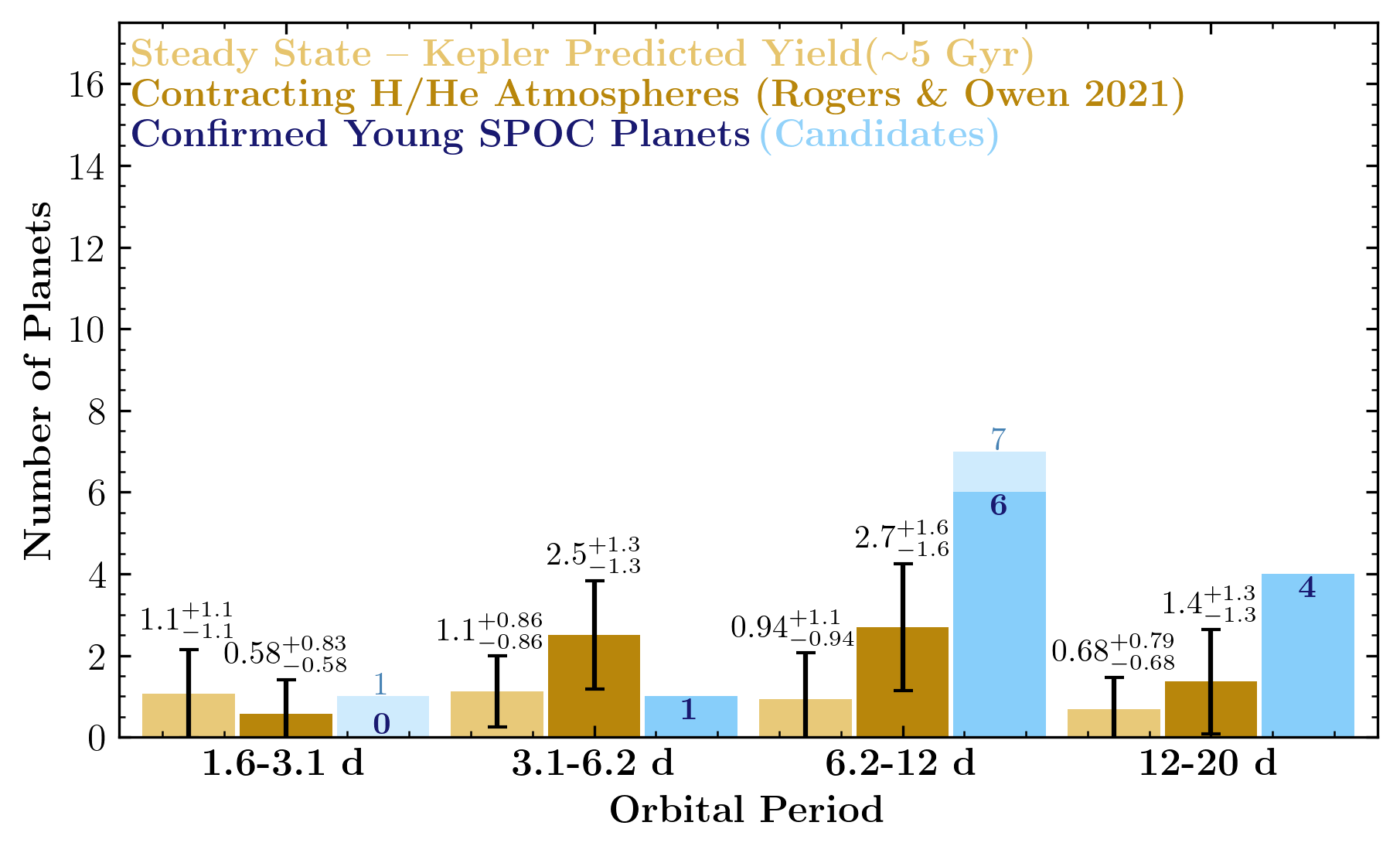}
    \includegraphics[width=0.49\linewidth]{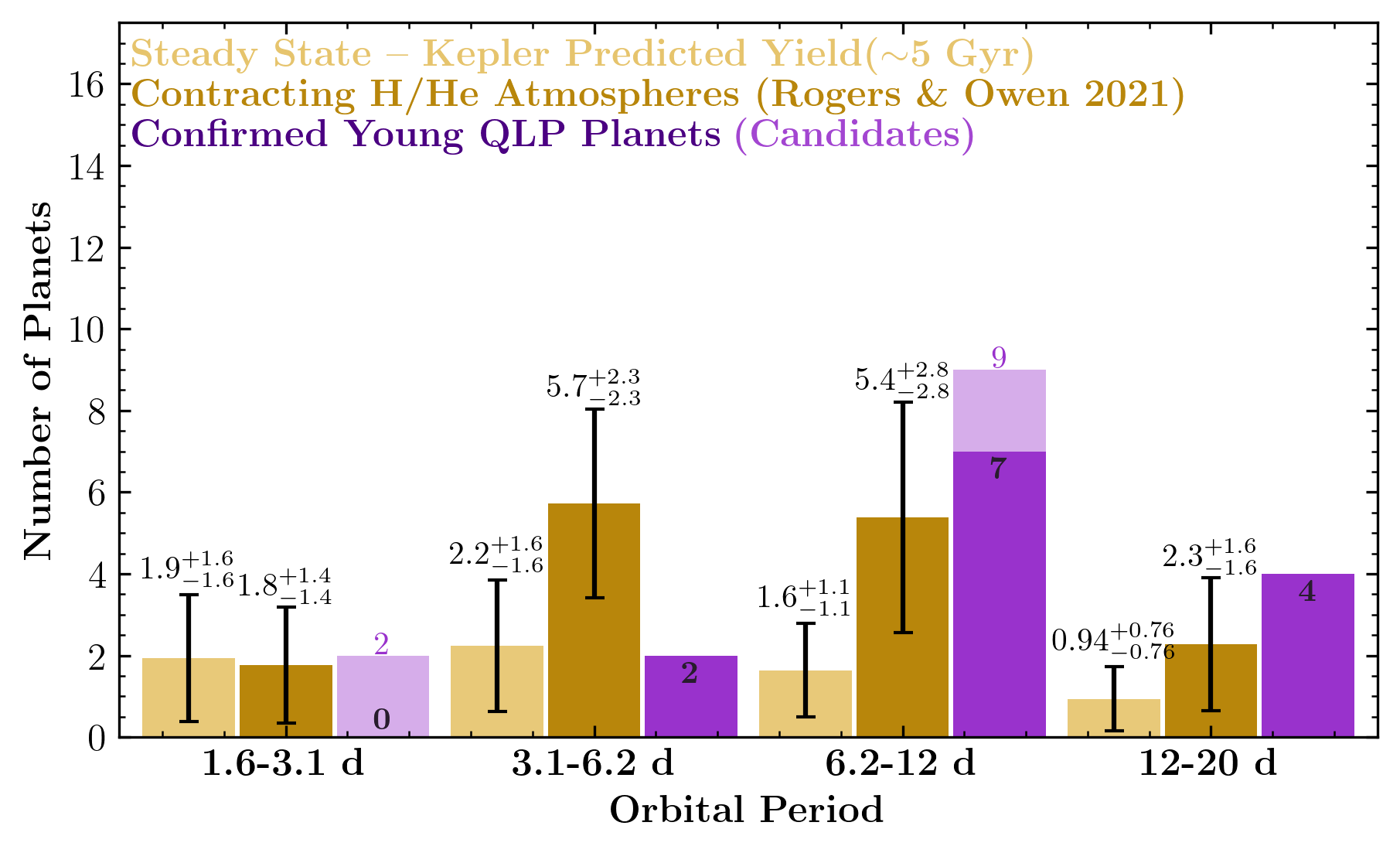}\\
        \includegraphics[width=0.49\linewidth]{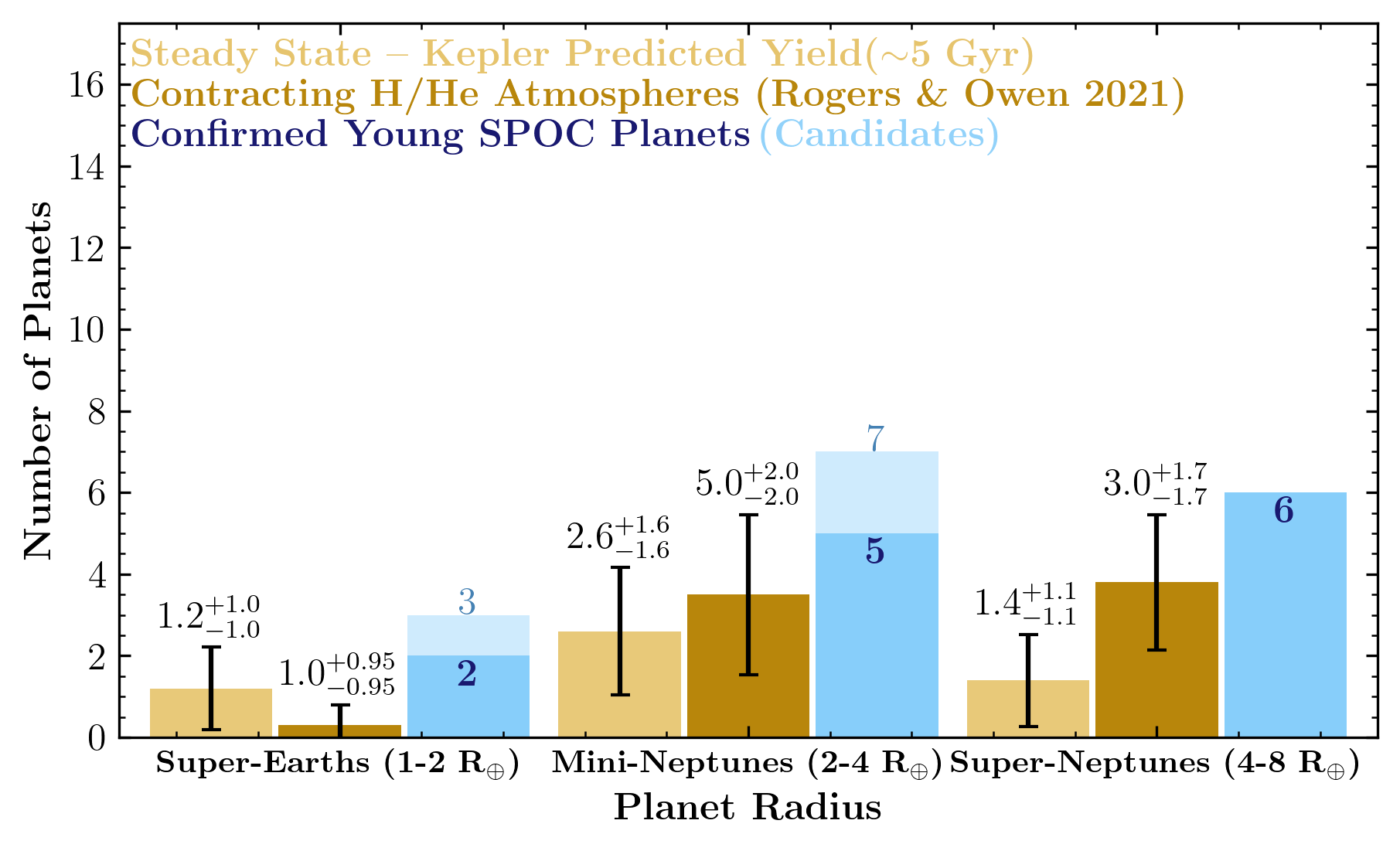}
    \includegraphics[width=0.49\linewidth]{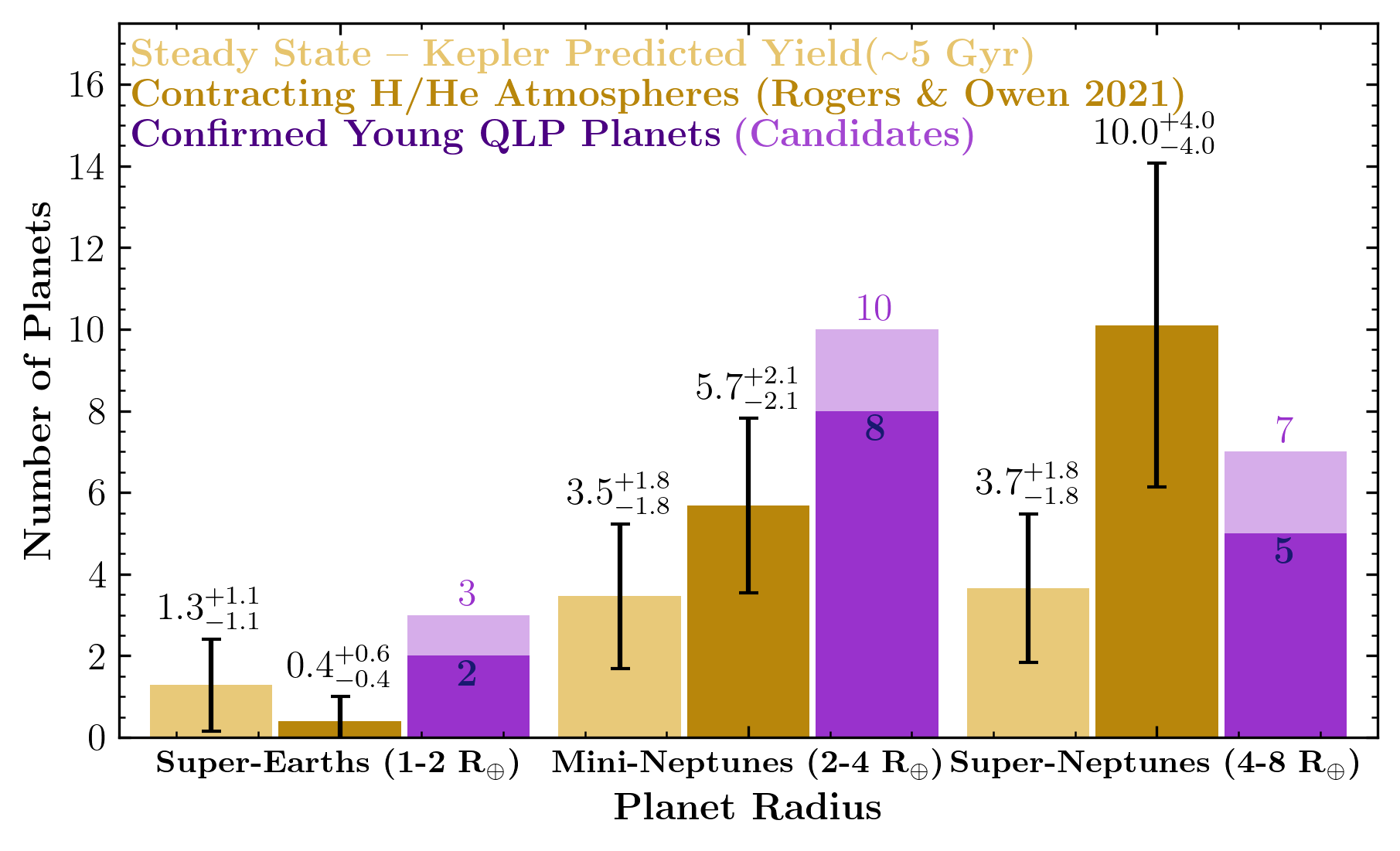}
    \caption{Number of combined confirmed planets (dark) and planet candidates (light) detected in our SPOC (left) and QLP (right) planet surveys compared to the expected yield (light yellow) from our forward model calculations using the Kepler occurrence rates \citep{Kunimoto2020} and the expected yield from evolution models for hydrogen-dominated (dark yellow) planets \citep{Rogers:2021} as a function of orbital period (top) and planet radius (bottom). We find a significant excess in planets with orbital periods between 6.2 and 12 days around young stars. We also find a mild excess of young super-Neptunes compared to the predicted Kepler yield. Note, the forward modeled population as a function of orbital period does not include planets in the super-Earth bin to allow for a more direct comparison to our estimated occurrence rates, as we do not attempt to derive an occurrence rate for the $1-2$\,$R_\oplus$ bin. 
   We find that the yield in both our QLP and SPOC surveys is consistent with evolutionary models for the contraction of hydrogen-dominated atmospheres undergoing photoevaporative escape.
    }
    \label{fig:forwardmodel}
\end{figure*}

We adopt the \citet{Kunimoto2020} Kepler occurrence rates for F, G, and K stars. We draw the occurrence rates assuming half-Gaussians based on the asymmetric uncertainties in \citet{Kunimoto2020} for each of the simulations. For each star in our parent sample, we simulate a planetary system based on its Kepler period-radius planet frequency grid for its specific stellar type.\footnote{The definition of the stellar-type boundaries are adopted to be the same as \citet{Kunimoto2020} For the M dwarfs in our sample, we use the occurrence rates for K stars to approximate their planet occurrence rate.}. The planets are first randomly assigned into a period-radius cell based on the two-dimensional cumulative probability. Then, their period and radius are drawn within the cell assuming log-uniform distributions in both axes. We then account for transit probability and detection probability as per the injection and recovery exercise described in Section~\ref{sec:injection}. For each planet, the transit probability is calculated as $0.8\times R_\star / a$, as we reject near-grazing candidates with impact parameters of $b>0.8$ due to their difficulty for confirmation. Detection probability is computed as the percentage of recovered planets within a single period-radius bin as shown in Figure~\ref{fig:recovery_maps} from our injection and recovery exercise. This forward modeling process is performed 100 times each for the QLP and SPOC samples to quantify its associated uncertainties. 
The resulting comparison between the expected population and that identified from the TESS sample is presented in Figure~\ref{fig:forwardmodel}.


We compare the predicted planet yield from the forward modeling to the detected sample further in Section~\ref{sec:fwdmod_comp}. The forward modeling comparison accounts for intrinsic differences in occurrence rates due to the differing host star populations sampled by the QLP and SPOC samples from TESS and the Kepler sample. Such differences, such as that of SPOC preferentially sampling later-type stars, may contribute to an intrinsic shift in the planet occurrence rates compared to that from the Kepler sample, as well as a difference in the measured occurrence rates between the two TESS subsamples.

\subsection{Approximate Bayesian Computation (ABC)}\label{sec:cosmoabc}

We use a population Monte Carlo approximate Bayesian computation \citep[ABC-PMC;][]{Beaumont:2009} method to estimate the occurrence rates of young planets. We use the ABC-PMC algorithm implemented in the python sampler \texttt{cosmoabc} \citep{cosmoabc}, described in \citet{Kunimoto2020}, which requires three elements: 
\begin{enumerate}
    \item a prior probability distribution, where we adopt the per-bin polynomial distribution as implemented in \citet{Kunimoto2020},
    \item a forward modeled population, used to generate a simulated set of data to compare to observations, and
    \item a distance function, used to determine the consensus between the simulated data and the observations.
\end{enumerate}

We use the prior probability distribution described in \citet{Kunimoto2020} and modify their forward modeling method and distance function to accurately represent the characteristics of the TESS data, rather than Kepler data. 
Typically, in Kepler occurrence studies, the recoverability of transit signals is computed by a function dependent on their estimated S/Ns based on the injection and recovery experiments \citep[e.g.][]{Christiansen:2015}.  In our study, we also require recoverability to be a function of the host star spectral type, such that the differences in stellar variability and transit duration are accounted for. 
This approach is also computationally more efficient compared to the per-star recovery map method detailed in our section~\ref{sec:forwardmodel}. We use one realization of an ABC-PMC run to test that the S/N recovery curve approach yields similar occurrence rates compared to the per-star recovery map approach and use the recovery curves for the final occurrence rate results presented in section~\ref{sec:discussion}.

We follow \citet{kovacs:2002} to derive the expected S/N of an injected transit signal for the SPOC observations. We adopt a similar S/N calculation for the QLP light curves, modified to account for its varying cadences (30 and 10 minutes),\footnote{30 and 10 minute observations; at the time we downloaded the light curves there were no 200 s FFI light curves made available on MAST} and we modify the calculation as follows. 

For each injected planet with transit depth $\delta$, we split the light curves of the star into two segments with the same cadence, and estimate the number of transits, $N_{tr,i}$,  and the average number of points in transit, $N_{in,i}$ for each segment.  The expected S/N is then expressed as:

\begin{equation}
    \mathrm{S/N_{\star}} = \delta\sqrt{\sum_i \frac{N_{tr,i}N_{in,i}}{\sigma_{i}^2}}, 
\end{equation}
where $\sigma_i$ is the per-point noise level, and is calculated as $1.48\times$ the median average deviation of the light curve.  

\begin{figure}
    \includegraphics[width=0.95\linewidth]{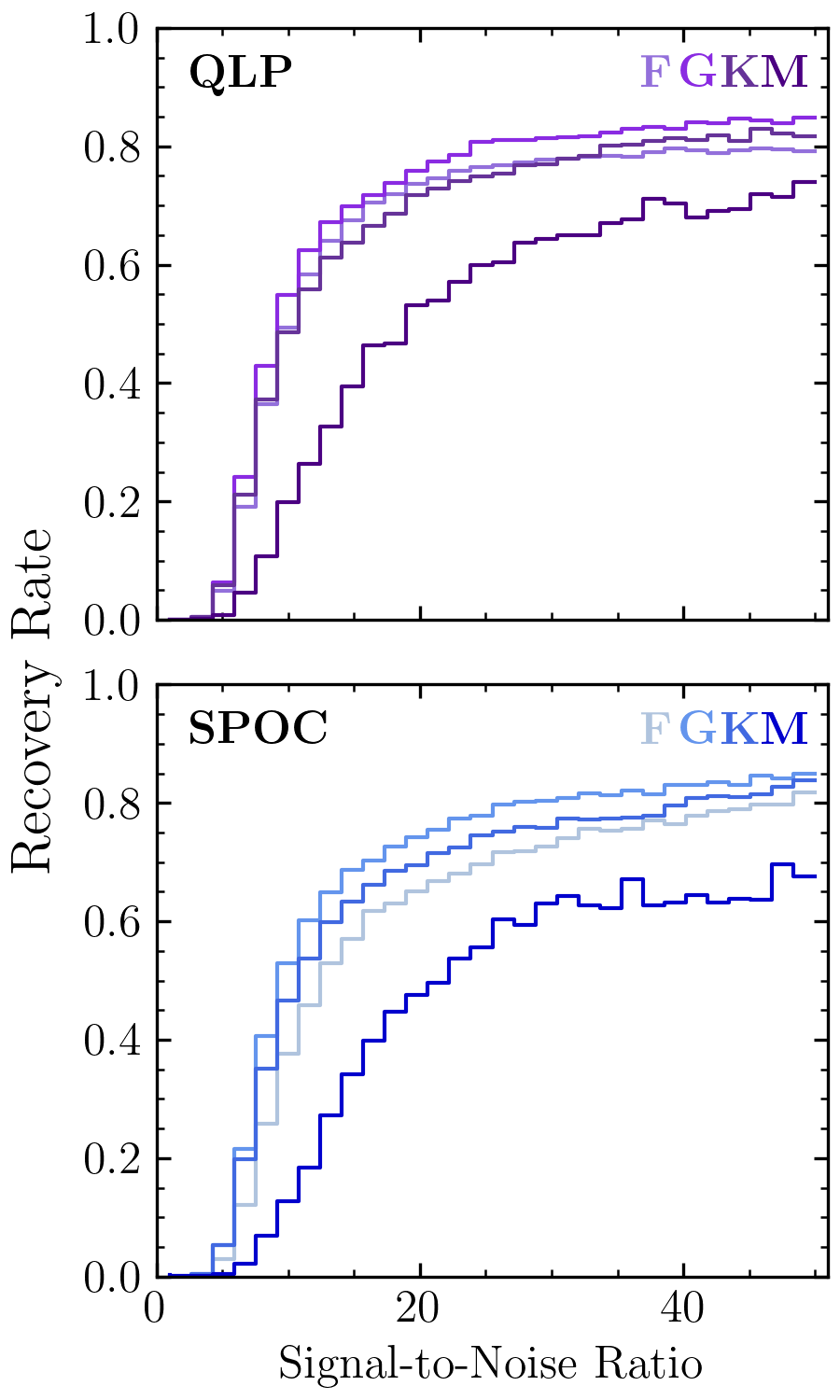}
    \caption{Recovery rate as a function of expected transit S/N derived for the QLP (top panel) and SPOC (bottom panel) samples. We compute the recovery rate for F-, G-, K-, and M-type stars individually to illustrate potential differences that might incur from differing levels of stellar variability between the spectral classes.}
    \label{fig:snr}
\end{figure}

Figure~\ref{fig:snr} shows our computed recoverability as a function of S/N by stellar type in the QLP and SPOC survey. Our recoverability for both surveys plateaus at $\mysim80\%$. 

In addition, we adopt a distance function similar to \citet{Kunimoto2020} Equation 21 but require the computation to be done over the radius, period, and stellar-type bins, rather than just radius and period. This allows us to account for the differences in stellar types in our parent stellar samples.

Following \citet{Hsu:2019} and \citet{Kunimoto2020}, we first fit for the occurrence rates over a fine period-radius grid. We break the period-radius space into bins between periods of 0.78, 1.56, 3.125, 6.25, 12.5, and 20 days and radii of 0.5, 0.71, 1.41, 2.0, 2.83, 4.0, 5.66, 8.0, 11.31, and 16\,$R_\oplus$. For each period-radius cell, the occurrence rate is obtained from fitting five radius bins (two on each side in radius dimension) simultaneously. The edge radius bins ($0.5-1.41\,R_\oplus$, $8-16\,R_\oplus$) are then excluded from the reported occurrence rates. These finer-resolution occurrence rate values are then combined into the final occurrence rates we report in larger cells in section~\ref{sec:discussion}.  

To account for the radius uncertainty of each planet and false-positive probability of the unconfirmed planet candidates in our sample, we perform multiple realizations of the occurrence rate calculations. At each realization, the radius of each planet is drawn from a Gaussian distribution about their fitted mean and standard deviation values. Each unconfirmed planet candidate is included, or excluded, based on a random draw as per their false-positive probability (see Section~\ref{sec:fprate}). 10 realizations of the SPOC and QLP samples are performed, with their resulting posteriors concatenated to attain the final occurrence rate posterior distribution, such that the uncertainties associated with planet radius and false positive rate are incorporated into our reported occurrence rates.

Additionally, we account for the possibility of stars falsely identified as members of kinematically young comoving groups in our parent stellar population. Our survey uses catalogs collated from Gaia kinematic associations described in \citet{Gagne:2018}, \citet{Kounkel:2019}, \citet{Ujjwal2020}, and \citet{Moranta:2022}. Our analysis of the rotational variability of the stars in our parent sample indicates $10\%-30\%$ of the stellar population does not exhibit rotational variations consistent with the literature ages. To incorporate this uncertainty in our parent stellar population, we randomly sample $70\%-90\%$ of the parent stellar population in each of our 10 realizations to estimate our occurrence rates. Each of the identified planet candidates and confirmed planets show signs of variability consistent with literature ages; as such we do not resample the planet population.

Due to our visual vetting process, we induce nonreproducible biases, especially near the detection limit, as signals induced by smaller planets are more challenging to identify by eye. This would manifest as an underrepresentation of the derived occurrence rates for small planets. In this work, we do not correct our completeness maps to account for the possibility of planets being mislabeled as false positives or false alarms in our vetting process. We therefore do not attempt to derive an occurrence rate for the $1-2\,R_\oplus$ bin, which would be most heavily impacted by this bias.

\subsection{Comparing Occurrence Rates and Forward Modeling Expectations} \label{sec:fwdmod_comp}

We find general agreement between the forward modeling results and the occurrence rate calculations. The results from our forward model exercises are presented in Figure~\ref{fig:forwardmodel}. 

Comparing bulk planet occurrence rates between different host star populations can be made difficult by the numerous factors on which planet frequencies depend. A number of works have demonstrated that the occurrence rate of small planets depends strongly on spectral type \citep[e.g.][]{Kunimoto2020}. The forward modeling approach accounts for spectral type dependencies of the host population. 

The forward modeled planet yield agrees well with that recovered from the young planet population for mini-Neptunes, and for short-period ($1.6-6.2$\,day) planets. This agreement is also reflected in the occurrence rate results, where the differences between our survey results and those from the Kepler sample differ by $<2\sigma$. We find an excess of super-Neptunes, and planets at longer period orbits compared to that expected from the forward model, resulting in a noteably higher occurrence rate for these subpopulations as discussed in Section~\ref{sec:mainresult}.

\subsection{Comparing the QLP and SPOC Results} 

The derived occurrence rates for our SPOC and QLP samples are presented in Figure~\ref{fig:occurrence_rates}. There exist differences between the QLP and SPOC samples at the $1-1.5\,\sigma$ level. We find a generally lower planet occurrence rate in the QLP sample compared to the SPOC sample. The small planet and parent stellar population sample size makes it difficult to interpret the validity of this difference. 

Figure~\ref{fig:stellar_pop_hist} shows the QLP parent sample is skewed toward earlier-type stars. \citet{Kunimoto2020} find K dwarfs to host $2-3\times$ more planets in the $2-8\,R_\oplus$ range compared to F dwarfs. The potential differences in the QLP and SPOC occurrence rates highlight the difficulty in a direct interpretation of planet frequency results without also accounting for the bulk parent stellar type. If such stellar-type dependencies are validated, it points to the paucity of small planets around F stars being developed early in their evolutionary history, as expected due to the differing levels of irradiation received by these planets around early-type stars. 

Target stars selected as a part of the initial TESS mission and TESS Guest Investigator Programs to have their light curves sampled at 2 minute cadences may bias the derived SPOC occurrence rate. These young stars may have been known to host planets (i.e., V1298 Tau) or have been identified as planet candidate hosts in the FFIs subsequently receiving SPOC observations during the extended TESS mission, resulting in an overall higher occurrence rate compared to the QLP sample. Further, the SPOC sample favors later-type stars, which are known to have higher occurrence rates \citep[e.g.][]{Kunimoto2020}. 

It is also likely the contamination rate for cluster membership of the parent population is different for QLP and SPOC. The low occurrence rate of planets in the QLP sample is more consistent with that of the mature planet population. To investigate the impact of our false-positive rate for cluster membership on the derived occurrence rates, we repeat our occurrence rate calculations as described in section~\ref{sec:or} without resampling the parent stellar population for the $10-30\%$ assumed contamination rate. Using the full stellar population observed by SPOC, we derive an occurrence rate of 28$_{-8.5}^{+11}$\% for mini-Neptunes and  25$_{-7}^{+8.6}$\% for super-Neptunes. For the full QLP population, we calculate an occurrence rate of $17^{+6.4}_{-5.0}\%$ for mini-Neptunes and  $8^{+2.9}_{-2.3}\%$ for super-Neptunes. The excess of super-Neptunes is still prevalent in these calculations; indicating that the uncertainties on our occurrence rates (see section~\ref{sec:mainresult}) are derived from the small planet number statistics rather than the membership contamination rate of our parent sample. Similarly, we find the excess of Neptune-sized planets with periods of $6.2-12$ day as well in the full stellar sample. We derive an occurrence rate of 17$_{-5.6}^{+7.3}$\% for the $6.2-12$ day bin with the SPOC sample and $8^{+3.5}_{-2.6}\%$ with QLP. Secure young star membership lists may help future studies resolve the discrepancies between the derived SPOC and QLP occurrence rates.

\section{Occurrence rate results}\label{sec:discussion} 

\begin{figure*}[ht!]
    \includegraphics[width=0.49\linewidth]{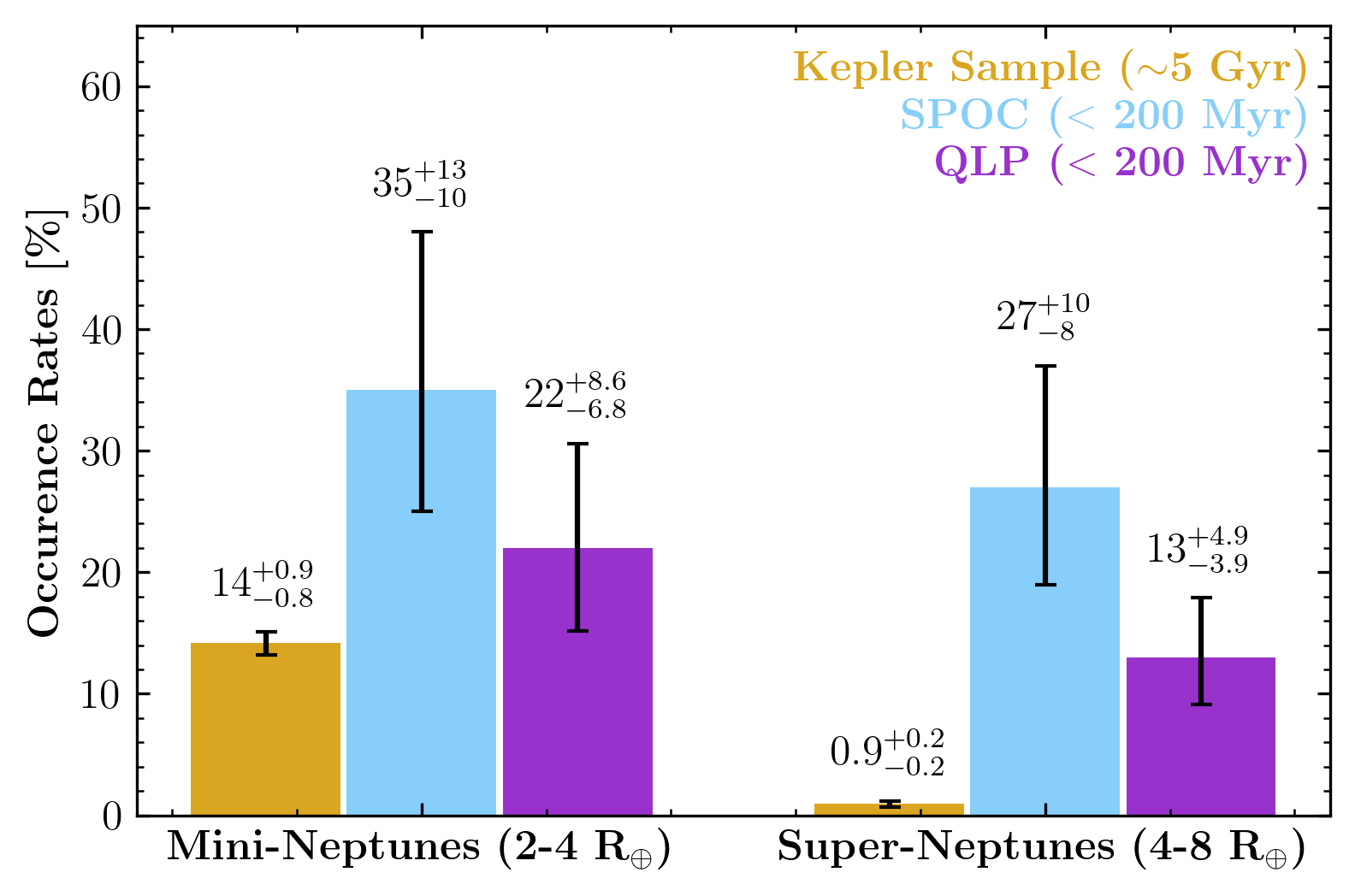}
    \includegraphics[width=0.49\linewidth]{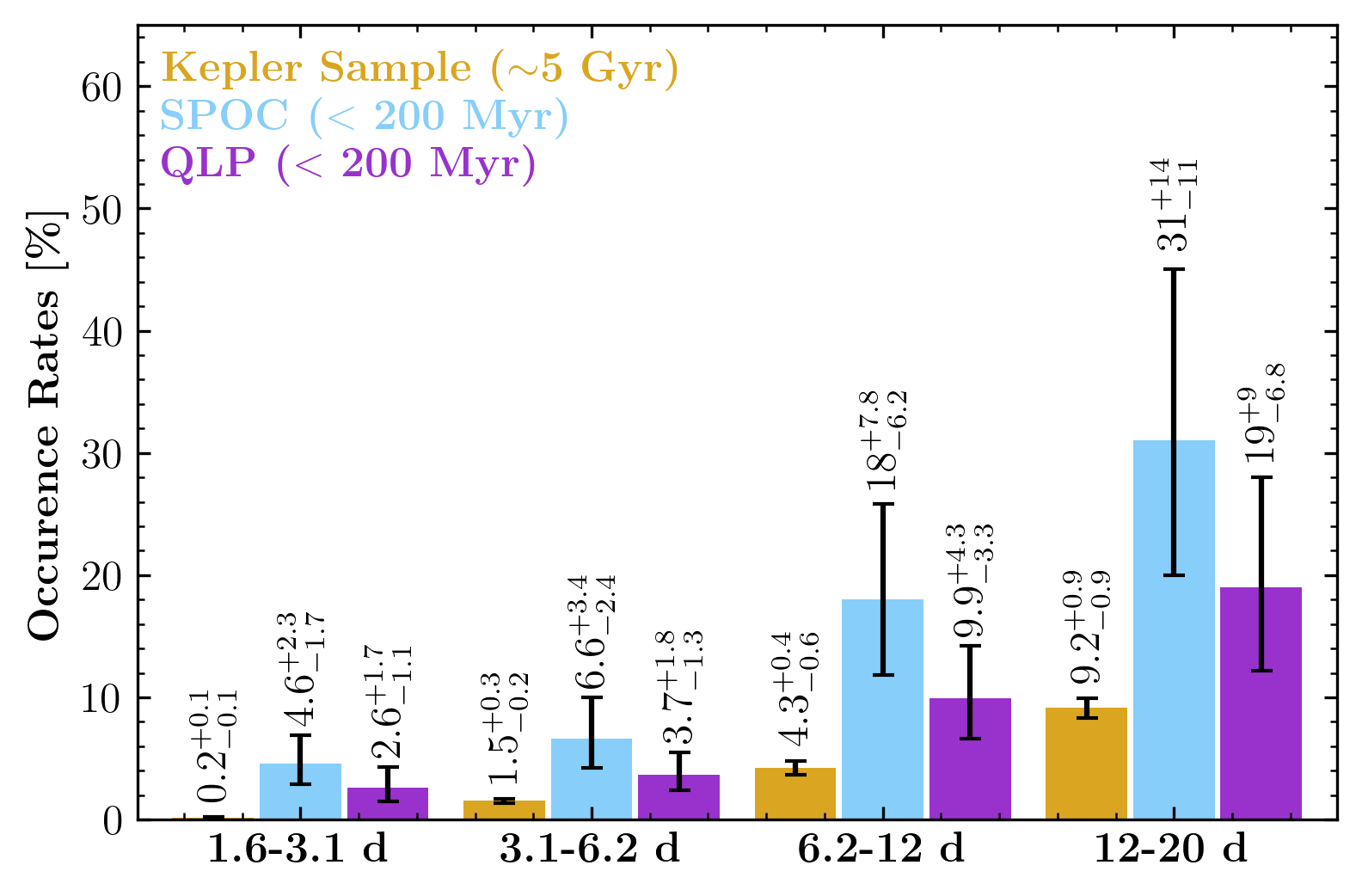}
    \caption{Occurrence rates for short-period, small planets around stars younger than 200 Myr in known clusters from the TESS 2 minute cadence observations (blue) and FFIs (purple) as a function of planet radius (left) and orbital period (right). We compare our derived occurrence rates to that of the mature population (gold) as derived from the Kepler yield \citep{Kunimoto2020}.}
    \label{fig:occurrence_rates}
\end{figure*}

The computed occurrence rates are shown in Figure~\ref{fig:occurrence_rates}. We break down the occurrence rates as a function of planet radius over the mini-Neptune ($2-4\,R_\oplus$) and super-Neptune ($4-8\,R_\oplus$) ranges. Due to only detecting two super-Earths in our SPOC and QLP surveys, we do not attempt to derive an occurrence rate for the $1-2\,R_\oplus$ bin. 
We also break down the occurrence rates as a function of orbital period, over the bins of $1.6-3.1$, $3.1-6.2$, $6.2-12$, and $12-20$ days. 

For short-period ($1.6-20$ days), small planets ($2-8\,R_\oplus$) younger than 200 Myr, we derive an occurrence rate of \spocor\ from our SPOC survey and \qlpor\ from our QLP survey. For comparison, the Kepler occurrence rate for similar planets in this period range is $15.0^{+1.1}_{-1.3}\%$. This result is qualitatively consistent with previous works that show an increase in the planet occurrence rate about young stars \citep[e.g.][]{Rizzuto:2017,Christiansen:2023,Fernandes:2023}. 

\subsection{The Planet Distribution about Young Stars is Different to that of the Mature Population} \label{sec:mainresult}
In both the QLP and SPOC samples, we find (1) a mild excess of super-Neptunes in the young planet population and (2) a significant excess of Neptune-sized planets at $\mysim10$ day orbital periods in the young planet population. 

\subsubsection{We Find a Mild Excess of Super-Neptunes Orbiting Young Stars}
Previous young planet censuses have found a higher occurrence rate of super-Neptunes \citep[e.g.][]{Rizzuto:2017, Christiansen:2023, Fernandes:2023} at ages up to that of the Hyades cluster. \citet{David:2021} and \citet{Berger:2023} found the distribution of small planets is continuously modified over a billion-year timescale, pointing to the role core-powered mechanisms play in shaping the mature planet population. 

Our work presents the radius distribution of planets amongst one of the youngest stellar distributions studied so far. We find a mild excess of super-Neptune-sized planets around the youngest surveyed population in the literature. The Kepler sample yielded an occurrence rate of $0.9\%\pm0.2\%$ for super-Neptunes in the mature population \citep{Kunimoto2020}. We derive higher occurrence rates for super-Neptunes (SPOC: \spocsn, differing to the equivalent Kepler occurrence rate at $2.9\sigma$ significance; QLP: \qlpsn, differing from Kepler at $2.8\sigma$) in our $<200$ Myr sample (see Figure~\ref{fig:occurrence_rates}). A similar significance exists in the forward modeling comparison, at $2.4\sigma$ for SPOC and $1.4\sigma$ for QLP. These qualitatively agree with previous works showing an excess of inflated planets around young stars.

\subsubsection{We Find a Significant Excess of $\mysim10$\,day Period Planets Around Young Stars}

The period distribution of young planets appears distinctively different from that of the mature population. Within our SPOC sample, we predict 0.94$^{+1.1}_{-0.94}$ planets between 6.2 and 12 days and find seven. The same excess of $\mysim10$\,day period planets is also apparent in our QLP sample. The Kepler statistic predicts 1.6$\pm1.1$ planets between 6.2 and 12 days, while we recover nine such planets.

This apparent period evolution between our $<200$ Myr sample and that of the Kepler distribution helps us understand the mechanisms that shape the sub-Saturn desert. \citet{Hallat:2022} and \citet{Thorngren:2023} find that XUV-driven mass loss can lead to runaway envelope stripping of the least dense hot Saturns. In this scenario, close-in planets with initial radii $>22\,R_\oplus$ undergo runaway mass loss within 100\,Myr. In addition, close-in planets that have undergone runaway mass loss remain puffy at the $\mysim 8\,R_\oplus$ radius due to residual heat, before cooling and contracting to $\mysim 4\,R_\oplus$ at time scales of over $\mysim500$ Myr. 

\citet{Hansen:2013} find that an initial overdensity of planets in the $5-10$\,day orbital period range is a natural result of the in-situ formation of rocky cores with the subsequent accretion of gaseous envelopes. This distance corresponds to the dust sublimation radius for Sun-like stars \citep{Swift:2013}. Tidal evolution models from \citet{Hansen:2015} show that this initial buildup of planets at $\mysim8$ days can migrate inward via tidal dissipation and eccentricity damping over a billion-year timescale if they exist in compact multiplanet systems that mutually undergo eccentricity excitation. Such tidally driven effects may be responsible for smoothing the overdensity of young planets into the mature planet period distribution.

\subsection{Comparison to Hydrogen-Helium-dominated Planet Evolution Models}\label{sec:rogers}

Mapping the distribution of young planets has the potential to disentangle otherwise degenerate modes of planet formation. Some super-Earths and Neptune-sized planets may have formed in close-in volatile- and gas-depleted regions of the protoplanet disk \citep[e.g.][]{Lee:2014}, while others may be composed of water-rich envelopes and were formed further out \citep[e.g.][]{Zeng:2019,Luque:2022}. These different formation modes produce planets of similar bulk physical properties in the mature planet distribution. In the water-rich scenario, the planets may have formed in the outer ice-rich protoplanet disk, migrating inward to their current locations \citep[e.g.][]{Mordasini:2009}. These planets may have silicate cores, icy mantles, and steam atmospheres \citep[e.g.][]{Rogers:2010,Zeng:2019}. 

In the alternate scenario, the planets may be born with a thick hydrogen-helium envelope surrounding an iron-silicate core in the volatile-depleted inner regions of the disk \citep[e.g.][]{Lee:2014}. Some of these hydrogen-helium-dominated planets may undergo runaway mass loss early in their evolution driven by boil-off \citep{Ginzburg:2016,Owen:2016}, photoevaporation \citep{Lopez:2013,owen:2013} and core-powered \citep{ginzburg2018, Gupta:2019} mechanisms. \citet{Owen:2023} suggested that a combination of these mechanisms is likely acting with timescales from $\mysim100$\,Myrs to gigayears, although the details of this timeline require further study. 

These different formation pathways differ most drastically early in their evolution. In particular, hydrogen-helium-dominated atmospheres are expected to experience rapid contraction within the first few hundred million to billion years of evolution \citep[e.g.][]{Lopez:2012,Chen:2016}, while steam-dominated atmospheres experience very mild radius contraction as they cool \citep{Lopez:2012,Lopez2017}, and our young planet population should resemble that of the Kepler distribution. Furthermore, atmospheric escape mechanisms should be evident with an increased frequency of planets $\gtrsim1.8\,R_\oplus$ around younger stars. As some of the planets have their atmospheres stripped and transition to below the radius valley, this frequency should decrease with time.

To test these predictions, we perform a forward modeling exercise using the models presented in \citet{Rogers:2021} and \citet{Rogers:2023}. These models consider the thermal contraction and photoevaporative mass loss for small, close-in exoplanets hosting hydrogen-helium atmospheres. We apply our forward modeling framework (Section~\ref{sec:forwardmodel}) to a population of synthesized planets about our stellar population, to test for the consistency between the observed distribution and a synthetic distribution of planets evolving through these physical processes.

For each star in the underlying SPOC and QLP stellar samples presented in Table~\ref{tab:stars}, we forward model 100 planets, assuming the underlying core mass, initial atmospheric mass fraction, core composition, and orbital period distributions inferred in \citet{Rogers:2021}. The synthesized planets have ages corresponding to that estimated for each respective association as listed in Table~\ref{tab:stars}. We then randomly draw planets from this sample to match the Kepler occurrence rates, as done in \citet{Rogers:2021}. This population, therefore, represents a sample of young planets about our observed stellar sample, under the photoevaporation model, which would eventually evolve to reproduce the Kepler sample of $\mysim5$~Gyr old super-Earths and mini-Neptunes.

We compare the predicted yield to our QLP and SPOC planet yields in Figure~\ref{fig:forwardmodel}. The models are in agreement at the $0.5-1.5\,\sigma$ level, showing the derived occurrence rates in this work are consistent with the contraction of hydrogen-dominated atmospheres undergoing photoevaporative escape \citep{Rogers:2021}. Of note, one can see that the occurrence rate of super-Neptunes in the \citet{Rogers:2021} models is increased when compared to the Kepler sample in Figure \ref{fig:occurrence_rates}. This is because the young mini-Neputunes in this model host significant hydrogen-helium atmospheres that are inflated to super-Neptune sizes for ages $<200$~Myrs. While the results are generally inconsistent with noncontracting steam-based atmospheres (see Figure \ref{fig:forwardmodel}), we leave model comparisons of evolving hydrogen vs. high mean-molecular weight (e.g. steam) atmospheres for future work. 

\section{Conclusion}\label{sec:conclusions}
Understanding how the occurrence rates of small, short-period planets evolve as a function of age can unveil the universal mechanisms driving planetary evolution. This paper presents the occurrence rates for young ($\leqslant200$ Myr), short-period $1.6-20$ days), small ($2-8\,R_\oplus$) planets from both the short-cadence and FFI light curves from TESS.

\begin{itemize}
    \item We conduct two independent planet surveys with the SPOC data (1927 stars) and the FFIs (7154 stars) from TESS targeting known moving groups and cluster members with literature ages $\leqslant$200 Myr \citep{Gagne:2018, Kounkel:2019, Ujjwal2020, Moranta:2022}.
    \item We identify a total of 23 planets, including four new planet candidates in the SPOC and QLP data previously not identified as TOIs. Initial ground-based follow up rule out any obvious astrophysical false positives.
    \item We derive occurrence rates for young, short-period, small planets using an Approximate Bayesian Computation approach, finding an increase in the occurrence of super-Neptunes within the $\leqslant200$ Myr population, consistent with models of planetary evolution, and a significant excess of planets residing at $\mysim10$ day orbital periods when compared to the mature population of exoplanets.
\end{itemize}

\section*{Acknowledgements}
We respectfully acknowledge the traditional custodians of the lands on which we conducted this research and throughout Australia. We recognize their continued cultural and spiritual connection to the land, waterways, cosmos, and community. We pay our deepest respects to all Elders, present and emerging, and the people of the Giabal, Jarowair, and Kambuwal nations, upon whose lands this research was conducted.

S.V. thanks the support of the Fulbright U.S. Student Program, which is sponsored by the U.S. Department of State and Australian-American Fulbright Commission. The contents of this paper are solely the responsibility of the author and do not necessarily represent the official views of the Fulbright Program, the Government of the United States, or the Australian-American Fulbright Commission.
G.Z. thanks the support of the Australian Research Council projects DE210101893 and FT230100517.
C.H. thanks the support of the ARC DECRA project DE200101840.

This research has made use of the NASA Exoplanet
Archive, which is operated by the California Institute of Technology,
under contract with the National Aeronautics and Space Administration
under the Exoplanet Exploration Program. 
Funding for the TESS mission is provided by NASA's Science Mission directorate. We acknowledge the use of public TESS Alert data from pipelines at the TESS Science Office and at the TESS Science Processing Operations Center. This research has made use of the Exoplanet Follow-up Observation Program (EXOFOP) website, which is operated by the California Institute of Technology, under contract with the National Aeronautics and Space Administration under the Exoplanet Exploration Program. This paper includes data collected by the TESS mission, which are publicly available from the Mikulski Archive for Space Telescopes (MAST).
Resources supporting this work were provided by the NASA High-End Computing (HEC) Program through the NASA Advanced Supercomputing (NAS) Division at Ames Research Center for the production of the SPOC data products.

\facilities{TESS.}

\software{\texttt{astropy} \citep{astropy, Astropy:2018, Astropy:2022}, \texttt{batman} \citep{batman}, \texttt{cosmoabc} \citep{cosmoabc}, \texttt{emcee} \citep{emcee}, \texttt{Matplotlib} \citep{matplotlib}, \texttt{Numpy} \citep{numpy}, \texttt{scipy} \citep{scipy}, and \texttt{TRICERATOPS} \citep{triceratopscode}.}

\appendix

\section{Inverted Light-curve Test}\label{sec:inverted}
False alarms are particularly prevalent at lower S/Ns. \citet{Kunimoto:2023} illustrate the presence of an excess of false alarms from BLS searches of TESS light curves at periods related to both astrophysical and instrumental signatures. Specifically, \citet{Kunimoto:2023} found buildups of BLS false-alarm detections at aliases of the TESS orbit, $\mysim13.7$ days. As this work performs a BLS search for planets with periods on the order of a TESS orbit and 0.5$\times$ a TESS orbit, we perform a blind planet search on both inverted and noninverted light curves to test the reliability of our detections.

We randomly sample 1000 stars in both our QLP and SPOC parent population to quantify our false-alarm rate. We ensure that both subsamples include the light curves of all planet hosts detected in this work (see Table~\ref{tab:tois}). For each individual light curve, we perform a random draw to determine if the light curve is to be inverted. For stars selected to have their light curves inverted, we follow \citet{Thompson:2018}, multiplying the zero-mean flux by $-1$ to invert the light curve. Each light curve is run through our planet detection pipeline (see Section~\ref{sec:pipeline}).

From our test, no signals from our inverted light-curve sample were wrongly classified as planet candidates. Of the positive light curves, all planet candidates were successfully recovered. In the QLP test sample, all the planet hosts were randomly selected to have their light curves inverted, none of which triggered a TCE. In the SPOC test sample, we identified TOI-2550.01 and DS Tuc Ab as planet candidates, neither of which had their light curves inverted. Additionally, HD 114082 b triggered a TCE,  however, the period and radius are not within our search parameters and therefore was not identified as a planet candidate. The remaining planet hosts all had their light curves inverted. The inverted light curve of AU Mic triggered a TCE, but our vetting procedure rejected the event due to asymmetries in the detected possible transit events.

\section{Robustness of the Occurrence Rate on Sample Assumptions}\label{sec:check}

To verify our derived occurrence rate calculations, we perform a series of checks to test the occurrence rates' robustness to astrophysical false positives and false positives induced by human vetting.

\begin{table}[]
\caption{Calculated Occurrence Rates}
\centering
\resizebox{\columnwidth}{!}{%
\begin{tabular}{@{}lcccc@{}}
\toprule \toprule
\multicolumn{1}{c}{Occurrence Rate Bin} &
  This work$^*$ &
  \begin{tabular}[c]{@{}c@{}}Without Filtering\\ for Binarity\end{tabular} &
  \begin{tabular}[c]{@{}c@{}}Confirmed \\planets only\end{tabular} &
  \begin{tabular}[c]{@{}c@{}}All planets and \\ planet candidates\end{tabular} \\ \midrule \midrule
\textbf{SPOC}                             &                       &  &                                      &                                      \\
Mini-Neputnes ($2-4\,R_\oplus$)  & 35$_{-10}^{+13}$\%    & 31$_{-9.4}^{+12}$\% (0.25$\sigma$)& 29$_{-9.3}^{+12}$\% (0.38$\sigma$)   & 36$_{-9.9}^{+12}$\% (0.063$\sigma$)    \\
Super-Neputnes ($4-8\,R_\oplus$) & 27$_{-8}^{+10}$\%  & 21$_{-6.3}^{+8.5}$\% (0.51$\sigma$) & 27$_{-8.5}^{+10}$\% (0$\sigma$)   & 31$_{-8.9}^{+12}$\% (0.29$\sigma$)   \\
$1.6-3.1$ days                            & 4.6$_{-1.7}^{+2.3}$\% & 3.6$_{-1.3}^{+1.9}$\% (0.39 $\sigma$) & \nodata & 4.4$_{-1.6}^{+2.2}$\% (0.072$\sigma$) \\
$3.1-6.2$ days                            & 6.6$_{-2.4}^{+3.4}$\% & 6.5$_{-2.5}^{+5}$\% (0.021$\sigma$) & 5.7$_{-2}^{+2.9}$\% (0.24$\sigma$) & 10$_{-3.3}^{+3.6}$\% (0.75$\sigma$)   \\
$6.2-12$ days                             & 18$_{-6.2}^{+7.8}$\%  & 18$_{-5.9}^{+6.7}$\%  (0$\sigma$)& 19$_{-6.4}^{+8.3}$\% (0.098$\sigma$)  &  21$_{-6.4}^{+8.4}$\% (0.29$\sigma$)   \\
$12-20$ days                              & 31$_{-11}^{+14}$\%  & 23$_{-8.8}^{+12}$\% (0.49$\sigma$) & 29$_{-10}^{+14}$\% (0.12$\sigma$)    & 31$_{-11}^{+14}$\% (0$\sigma$)       \\ 
\midrule 
\textbf{QLP}                             &                       &  &                                      &                                      \\
Mini-Neputnes ($2-4\,R_\oplus$)  & 22$_{-6.8}^{+8.6}$\%    & 18$_{-5.5}^{+7.9}$\% ($0.39\sigma$)  & 17$_{-5.1}^{+7.2}$\% (0.51$\sigma$)   & 19$_{-6.2}^{+8.1}$\% (0.28$\sigma$)    \\
Super-Neputnes ($4-8\,R_\oplus$) & 13$_{-3.9}^{+4.9}$\%     & 9.5$_{-2.9}^{+4}$\% (0.63$\sigma$) & 8.0$_{-2.3}^{+3.0}$\% (0.98$\sigma$)   & 10$_{-3.1}^{+4.1}$\% (0.53$\sigma$)   \\
$1.6-3.1$ days                            & 2.6$_{-1.1}^{+1.7}$\% & 1.6$_{-0.61}^{+0.84}$\% ($0.63\sigma$)  & \nodata & 2.0$_{-0.7}^{+1.1}$\% (0.36$\sigma$) \\
$3.1-6.2$ days                            & 3.7$_{-1.3}^{+1.8}$\% & 3.9$_{-1.3}^{+1.8}$\% ($0.090\sigma$) & 2.8$_{-1}^{+1.3}$\% (0.47$\sigma$) & 3.5$_{-1.3}^{+2}$\% (0.088$\sigma$)   \\
$6.2-12$ days                             & 9.9$_{-3.3}^{+4.3}$\%  & 7.8$_{-2.7}^{+3.8}$\% ($0.42\sigma$)& 7.3$_{-2.6}^{+3.4}$\% (0.53$\sigma$)  & 8.6$_{-2.8}^{+3.7}$\% (0.26$\sigma$)   \\
$12-20$ days                              & 19$_{-6.8}^{+9}$\%    & 14$_{-5.5}^{+7.6}$\% ($0.49\sigma$)& 13$_{-4.8}^{+6.9}$\% (0.61$\sigma$)    & 15$_{-5.6}^{+7.7}$\% (0.39$\sigma$)       \\ 
\bottomrule
\end{tabular}
}

\label{tab:my-table}
$^*$ Sigma differences are computed against this column. These values are the same as presented in Figure~\ref{fig:occurrence_rates}.
\end{table}

\subsection{Impact of the Renormalized Unit Weight Error Threshold on the Occurrence Rates}
We investigate the impact of binarity on our derived occurrence rates by repeating our occurrence rate calculations without removing high-RUWE stars. We collate our stellar parent population following Section~\ref{sec:stellar_pop}. We find 8852 stars with TESS observations. Of which, 8505 were observed in the FFIs and 2243 were observed in the target pixel stamps.

The occurrence rates derived from the full sample, without attempting to filter for binarity, agree with our presented occurrence rates at the $<1\sigma$ level. Similarly to the occurrence rates presented in Figure~\ref{fig:occurrence_rates}, we find a mild excess of super-Neptunes and a significant excess of Neptune-sized planets at orbital periods of $\mysim10$ days in the population without filtering for binarity. We search for transiting planets following Section~\ref{sec:pipeline}, and characterize our pipeline completeness following Section~\ref{sec:injection}. The planet sample recovered is identical to the sample presented in Table~\ref{tab:tois}. We follow Section~\ref{sec:or}, and perform 10 realizations of both the QLP and SPOC occurrence rate calculations, with each realization randomly sampling $70-90\%$ of the stellar population to account for the membership contamination rate. See Table~\ref{tab:my-table} full results.

\subsection{Impact of the False-positive Rate Assumptions on the Occurrence Rates}
To investigate the degree to which our adopted false-positive rate assumptions via \texttt{TRICERATOPS} impact our derived occurrence rates, we explore the bounding cases where (a) all planet candidates are counted as confirmed planets and (b) only verified/published planets are included in our occurrence rate calculation and all planet candidates are treated as false-alarms. We calculate the occurrence rates following Section~\ref{sec:or}. 

In both bounding cases, we find that the main conclusions presented in Section~\ref{sec:discussion} are also present: a mild excess of super-Neptunes and a significant excess of planet with orbital periods of $\mysim10$ days when compared to the Kepler statistics. Our full results are presented in Table~\ref{tab:my-table}.

\section{New Planetary Candidates}\label{newplanets}

We report from our planet search the following new planet candidates previously unidentified in the public TESS candidate list: TIC 150070085.01 (SPOC and QLP), TIC 88785435, and TIC 434398831.01 and TIC 434398831.02 (QLP).

All new candidates have been submitted to the TFOP WG for ground-based follow-up observations for further confirmation and characterization. We also conducted photometric and spectroscopic ground-based follow-up observations for TIC 434398831 to allow us to rule out any false-positive scenarios. Further, TIC 434398831 has been observed by the CHaracterising ExOPlanet Satellite \citep[CHEOPS;][]{Cheops}{}{} as a part of AO-4 program PR230002 (PI: S. Vach). 


Precise planet radii and periods are key for calculating planetary occurrence rates. To properly characterize the newly identified planet candidates, we perform a global model to constrain the planetary parameters for the individual systems. We follow \citet{Zhou2021} and model the stellar properties and transit light-curve parameters simultaneously. The free parameters for our global model are stellar mass, $M_\star$, stellar radius, $R_\star$, the orbital period, $P$, the time of transit center, $T_0$, the radius ratio, $R_P/R_\star$, and transit inclination $i$, and eccentricity parameters $\sqrt{e} \cos \omega$ and $\sqrt{e} \sin \omega$ that influence the transit duration. The limb-darkening parameters are fixed to those interpolated from \citet{Claret2017}. The transits were calculated using the \texttt{BATMAN} \citep{batman} implementation of the \citet{Mandel:2002} models. At each iteration, we interpolate the stellar parameters onto the MIST isochrone \citep{Dotter:2016} to estimate their magnitudes. These isochrone magnitudes are compared against that observed in the Gaia, \citep{Gaia2016}, the Two Micron All Sky Survey, and Wide-field Infrared Survey Explorer bands, corrected for the distance modulus via the Gaia parallax. We calculate a likelihood function incorporating the magnitudes and light curves. The best-fit parameters and posteriors are explored via a Markov Chain Monte Carlo run via \textsc{emcee} \citep{emcee}. The derived planet radii and associated uncertainties are incorporated into our occurrence rate calculations (See Section~\ref{sec:or}). We present our best-fit values for the new planet candidates identified in this work below.

\subsubsection*{TIC 150070085}
\begin{figure}[ht!]
    \centering
    \includegraphics[width=0.7\linewidth]{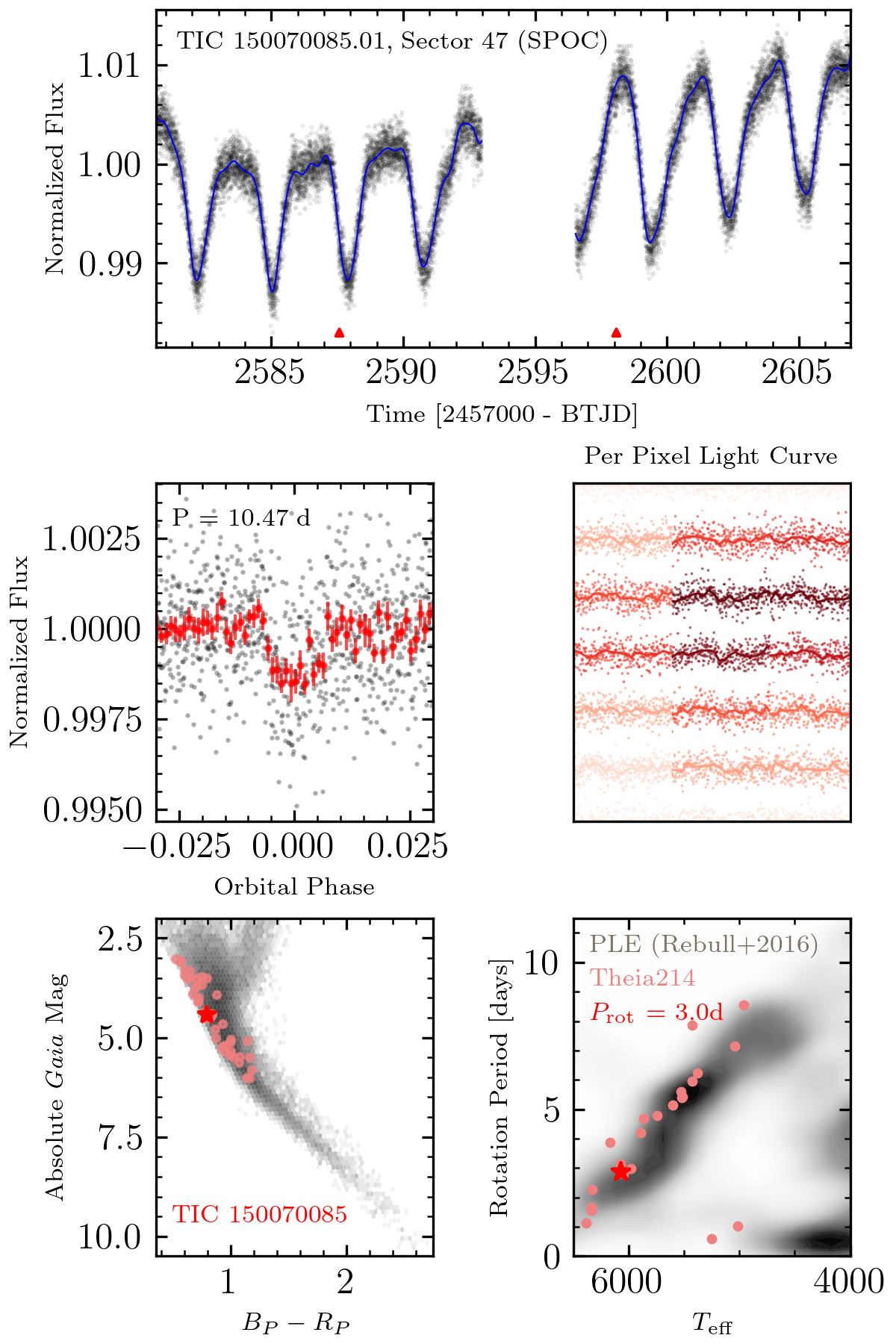}
    \caption{Vetting plot for TIC 150070085.01. The upper panel shows the TESS SPOC light curve (gray) from Sector 47. We model the stellar variability with a spline (blue) and mark the transits in the full light curves (red triangles). The middle panels show the phase-folded binned light curve (red) on the left using the BLS period and transit times overlaid with the full phase-folded FFI light curve (gray) and the per-pixel FFI light curves on the right. The bottom panel shows the CMD of the candidate (red star) and its comoving group (light red), Theia 214 (96 Myr). The Kepler Input Catalog \citep{Mathur:2017} CMD is overplotted for reference (gray-scale). On the right, we plot our derived stellar rotation periods as a function of effective temperature of TIC 150070085 (red star) and the members of Theia 214 (light red). The rotation periods of the 120 Myr old Pleiades association are overlaid in gray for reference \citep{Rebull:2016}}
    \label{fig:tic150070085}
\end{figure}

TIC 150070085 was observed by TESS in Sectors 20 and 47, receiving both 2 minute target pixel stamps and FFI observations. TIC 150070085.01 was identified in both our SPOC and QLP planet search with a signal-to-pink noise ratio of 8.2 and 9.5 respectively. Our vetting diagnostic figure for TIC 150070085.01 is shown in Figure~\ref{fig:tic150070085}. We utilize the 2 minute cadence observations for our global model. 

TIC 150070085 ($T_\mathrm{eff} = 6070\pm146$ K) is located in the Theia 214 moving group. We use our derived age to inform our model by constraining the stellar age. Our global best-fit model for the TIC 150070085 system finds $R_P = 3.64\pm0.41\,R_\oplus$, $t_0 = 2458843.8908\pm0.0041$ BJD, $P = 10.4745\pm 0.0036$ days, and $t_\mathrm{dur} = 0.1466\pm 0.0031$ days. The stellar rotation period, $P_\mathrm{rot} = 3.0$\,days, calculated from the QLP light curves shows the rotational spread for Theia 214 is consistent with the literature age of $\mysim100$ Myr.

\subsubsection*{TIC 88785435}
\begin{figure}[ht!]
    \centering
    \includegraphics[width=0.7\linewidth]{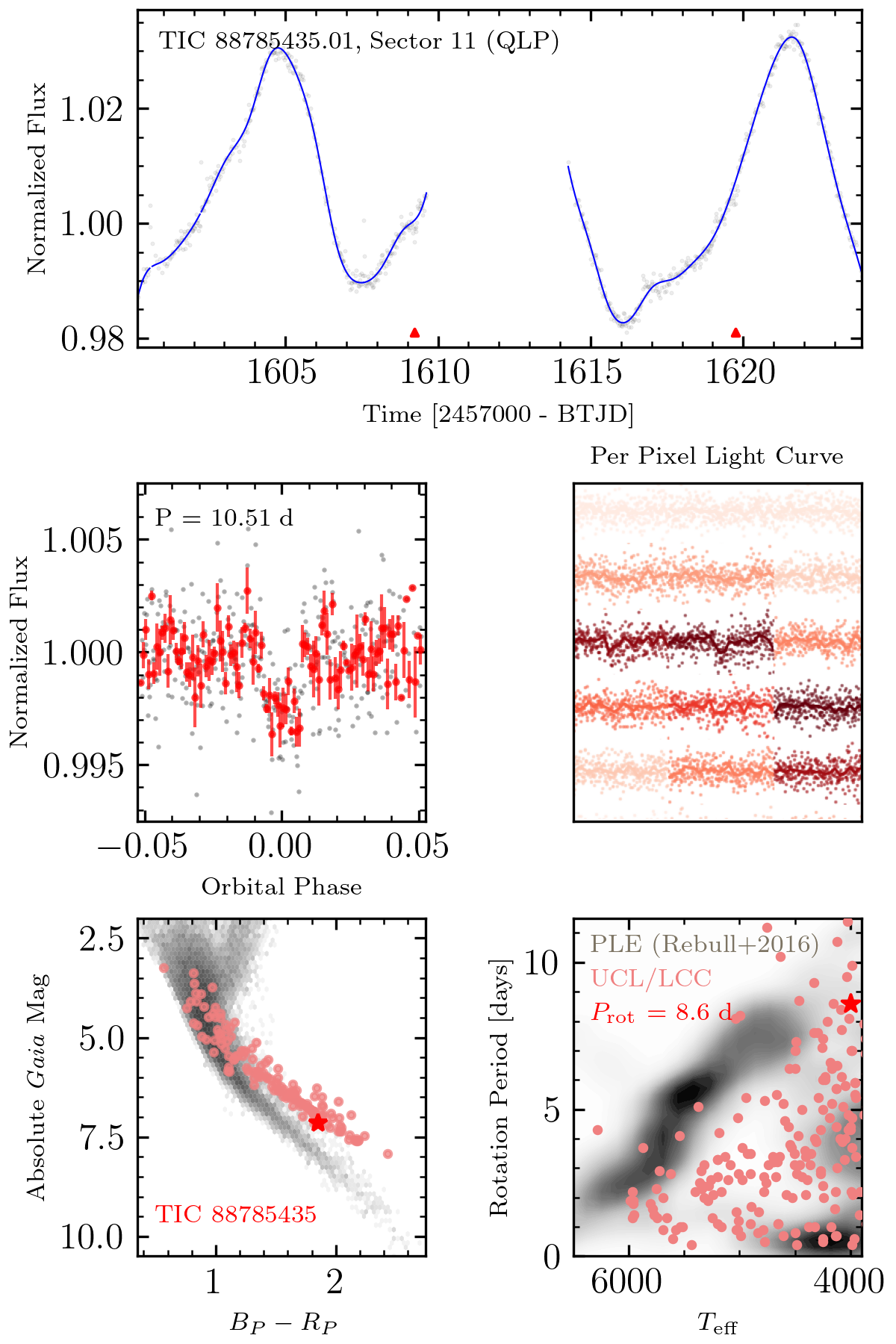}
    \caption{Vetting plot for TIC 88785435.01 following the style of Figure~\ref{fig:tic150070085}.}
    \label{fig:tic88785435}
\end{figure}

TIC 88785435 was observed by TESS in Sectors 11 and 38 in the FFIs. TIC 88785435.01 was identified in our QLP planet search with a signal-to-pink noise ratio of 9.47. Figure~\ref{fig:tic88785435} shows our vetting diagnostic figure for TIC 88785435.01.

TIC 88785435 ($T_\mathrm{eff} = 4000\pm150$ K) is located in the UCL/LCC moving group. We use our derived age to inform our model by constraining the stellar age. Our global best-fit model for the TIC 88785435 system finds $R_P = 4.88\pm0.28\,R_\oplus$, $t_0 = 2458609.2308\pm0.0036$ BJD, $P = 10.508907\pm 0.000035$ days, and $t_\mathrm{dur} = 0.1466\pm 0.0031$ days. The stellar rotation period, $P_\mathrm{rot} = 8.6$\,days, is calculated from the QLP light curves, consistent with the age of literature age of UCL/LCC.

\subsubsection*{TIC 434398831}\label{sec:4343}

TIC 434398831 (S. Vach et al. 2024, \textit{in preparation}) received FFI observations over TESS Sectors 6, 33, 43, 44, and 45. Our planet search identifies two planet candidates, TIC 434398831.01 (signal-to-pink noise ratio of 11.8) and TIC 434398831.02 (signal-to-pink noise ratio of 15.7). Our best-fit parameters for TIC 434398831.01 are $R_P = 3.65\pm0.28 \,R_\oplus$, $P = 3.68551266 \pm 0.000019$ days, $t_0 = 2458468.6318 \pm 0.0052$ BJD, and $t_\mathrm{dur} = 0.1061 \pm 0.0014$ days,  and for TIC 434398831.02 they are $R_P = 5.41\pm0.32\,R_\oplus$ $P = 6.210377\pm0.000017$ days, $t_0 = 2458470.6110 \pm 0.0027$ BJD, and $t_\mathrm{dur} = 0.1289 \pm 0.010$ days.

The host star TIC 434398831 is on the pre-main sequence, with an effective temperature of $T_\mathrm{eff} = 5554\pm 144$ K. Previous works place TIC 434398831 in the Theia 116  moving group. We measure the stellar rotation period, $P_\mathrm{rot} = 2.0$ days, using the FFI light curves. After analyzing the members of Theia 116, we find the rotational spread to be consistent with $\mysim50$ Myr, agreeing with the isochrone-derived age. 

Three transits of TIC 434398831.01 and two transits of TIC 434398831.02 have been observed by ESA's CHEOPS mission (AO-4 program PR230002). Paired with extensive LCO ground-based follow up of TIC 434398831.01 and .02, we were able to clear the field of NEBs and rule out any false-positive scenarios. Therefore, we classify TIC 434398831.01 and .02 as confirmed planets in our occurrence rate calculations and adopt false-positive rates of zero for both (see Table~\ref{tab:tois}).

\begin{figure*}[ht!]
    \centering
    \includegraphics[width=0.49\linewidth]{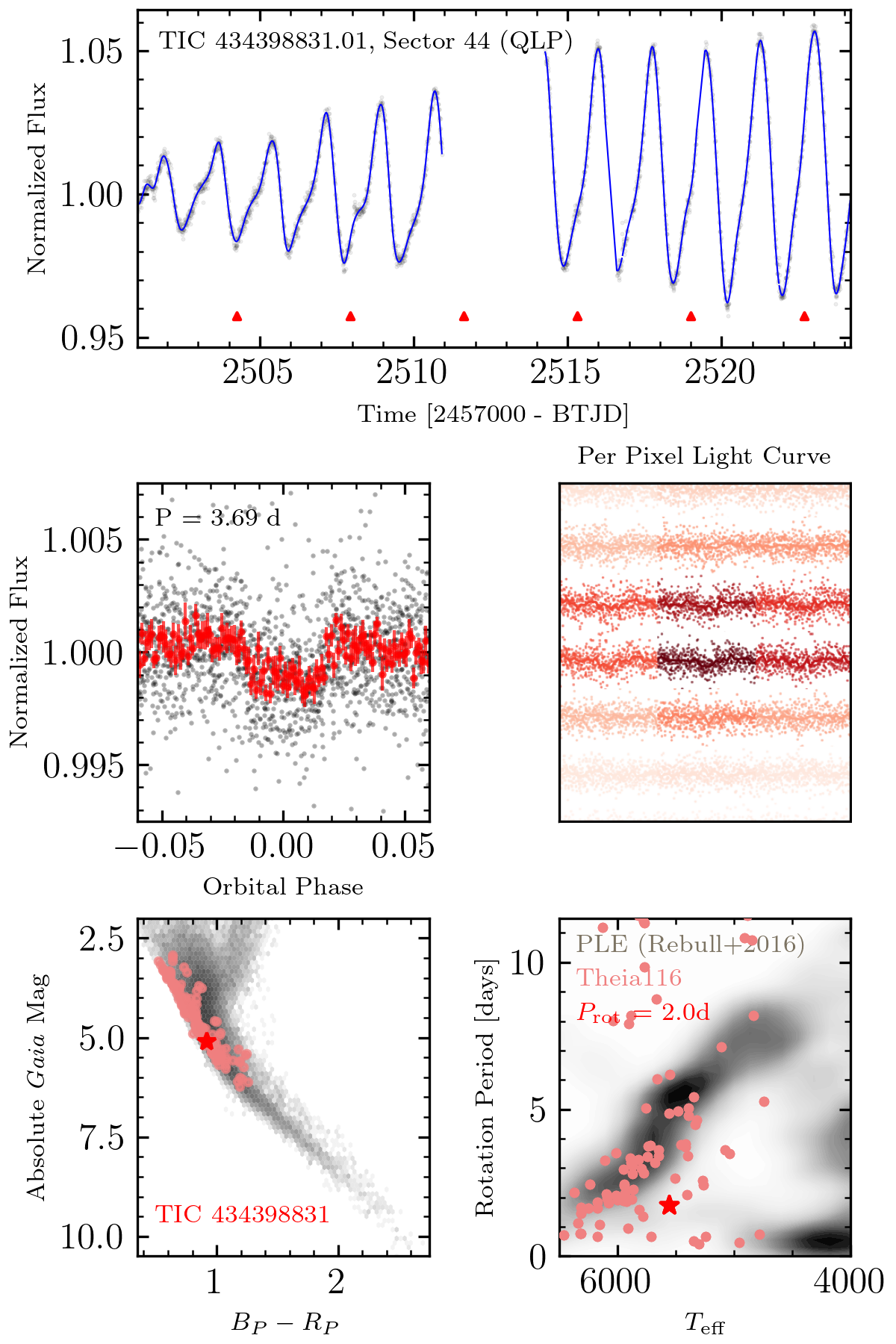}
        \includegraphics[width=0.49\linewidth]{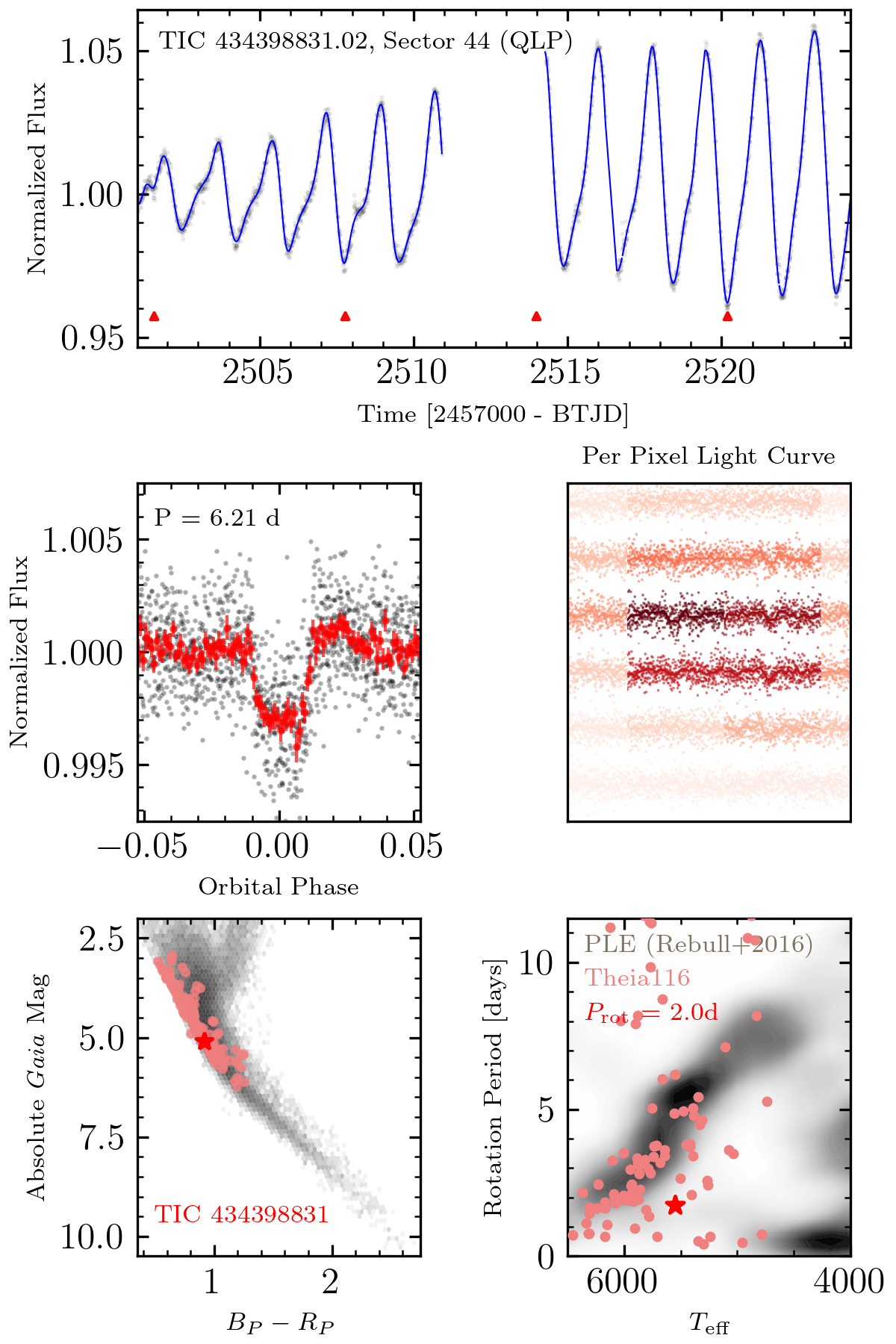}
    \caption{Vetting plots for TIC 434398831.01 (left) and TIC 434398831.02 (right) following the style of Figure~\ref{fig:tic150070085}.}
    \label{fig:tic434398831}
\end{figure*}

\section{Rotation Analysis Plots}\label{sec:rotation_plots}

\begin{figure}[ht!]
    \centering
    \begin{tabular}{l}
    \textbf{Example: Rotational variability}\\
    \includegraphics[width=0.75\linewidth]{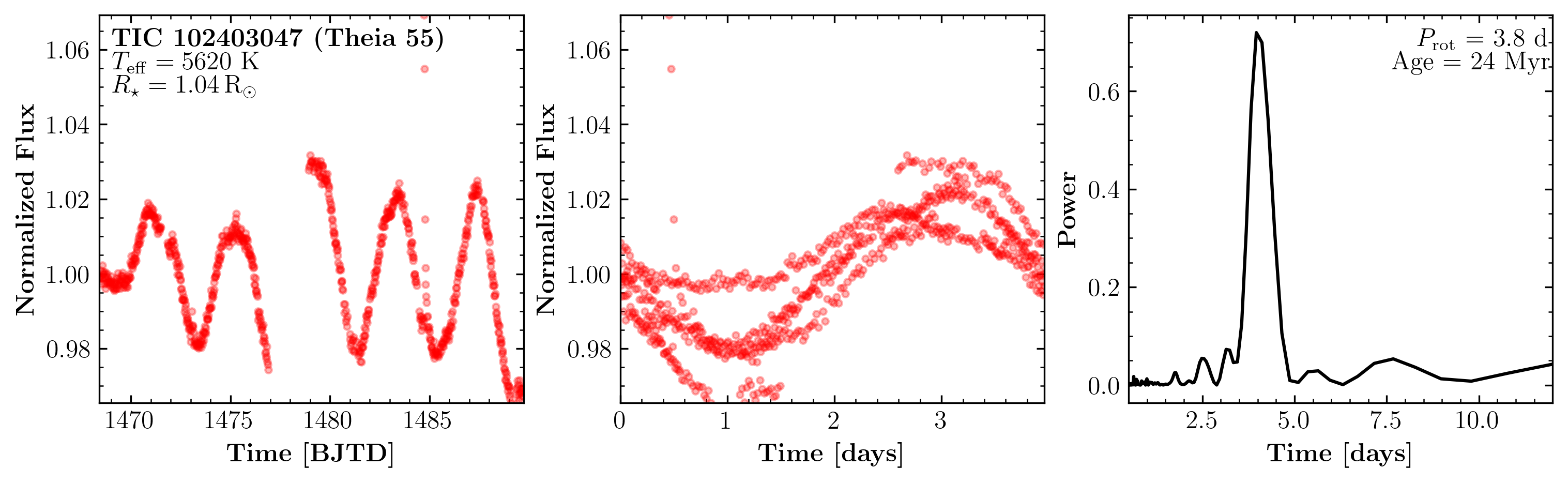} \\
    \textbf{Example: Non-rotational variability}\\
    \includegraphics[width=0.75\linewidth]{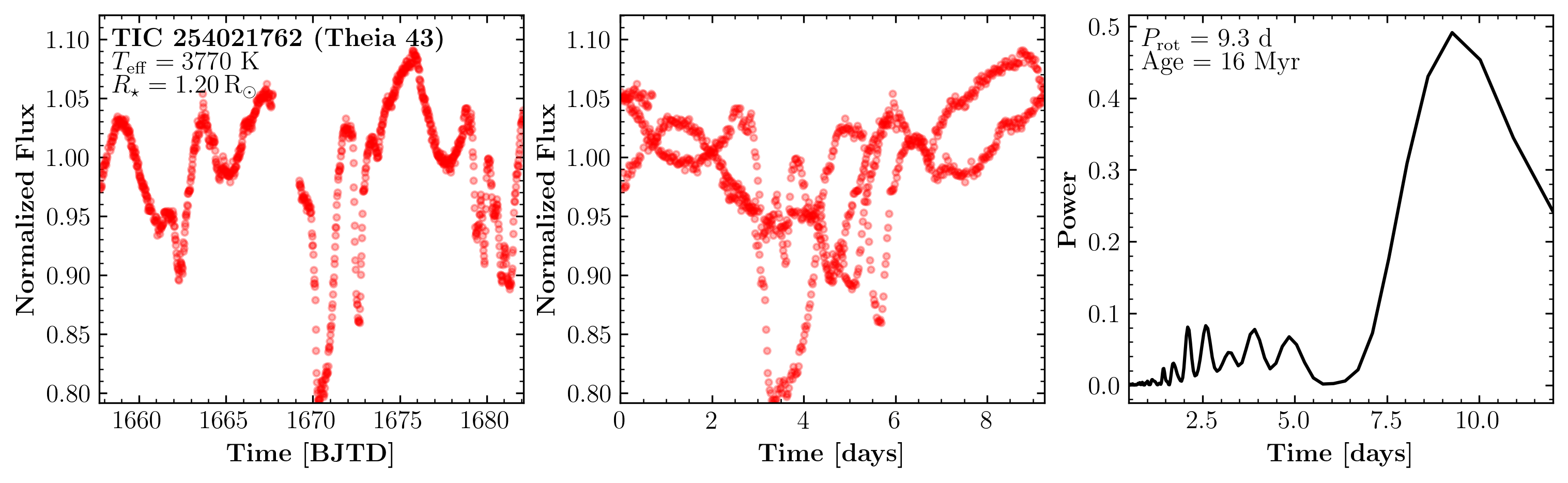} \\
    \textbf{Example: No variability}\\
    \includegraphics[width=0.75\linewidth]{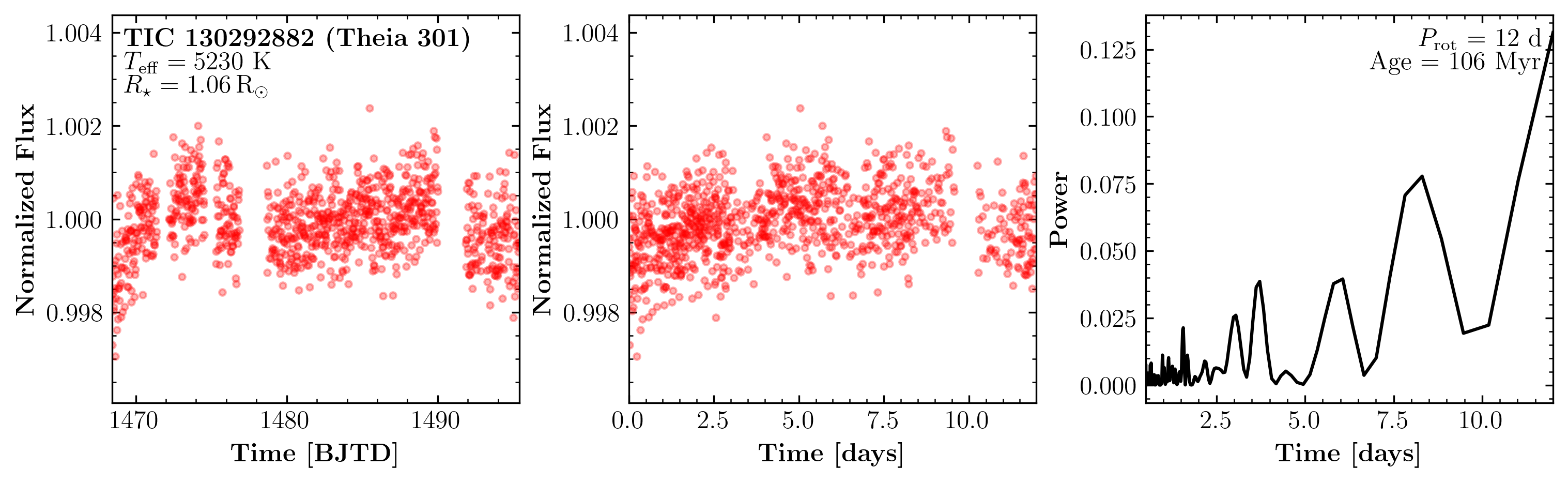} \\
    \textbf{Example: Binarity}\\
    \includegraphics[width=0.75\linewidth]{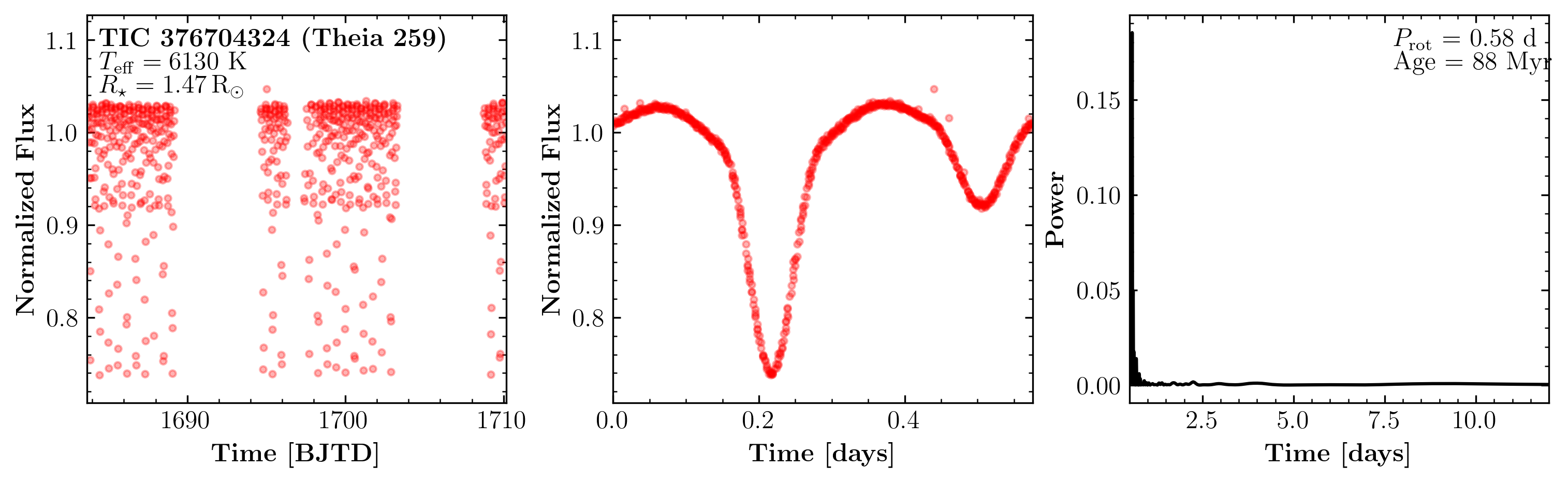} \\
    \end{tabular}
    \caption{Sample rotation vetting plots. Each row shows the nondetrended FFI light curve (left) from a single sector of TESS observations, the phase-folded light curve based on the measured rotation period (middle), and the Lomb-Scargle \citep{lomb1976, scargle1982} periodogram (right) sampling rotation periods between 0.5 and 12 days. The first panel displays an example rotational vetting plot in which a measured rotational period was identified. The second panel is an example of a variable star, but an incorrect/no rotation period is measured. The third panel illustrates an example where there is no apparent activity or rotation measured. The fourth panel is an example of a binary system.}
    \label{fig:rotation_examples}
\end{figure}

Our parent stellar population is a collation of multiple catalogs (see section~\ref{sec:stellar_pop}) from various clustering studies \citep{Gagne:2018,Kounkel:2019,Ujjwal2020, Moranta:2022}. To independently measure the field star contamination rate of our collated stellar population, we make use of the TESS FFI light curves to measure the rotation periods and variability signatures of stars in our survey with $T_\mathrm{eff} < 6500$ K. As we do not attempt to measure rotation periods longer than the duration of a single TESS orbit, we sample periods between 0.5 and 12 days using a Lomb-Scargle periodogram. Figure~\ref{fig:rotation_examples} shows examples of vetting figures that were used to classify rotational modulation light curves. We visually examine each light curve for signatures of rotational variability, and report the measured rotation periods where possible in Table~\ref{tab:stars}. 

\bibliography{refs}{}
\bibliographystyle{aasjournal}

\end{document}